\documentclass[preprint]{elsarticle}
\pdfoutput=1

\usepackage{lineno,hyperref}

\usepackage{graphicx}
\usepackage[super]{nth}
\usepackage{caption}
\usepackage{subcaption}
\usepackage{xcolor}
\usepackage{xurl}
\usepackage{array}
\usepackage{amssymb}
\usepackage{pifont}
\newcommand{\cmark}{\ding{51}}%

\usepackage{array}
\newcolumntype{P}[1]{>{\centering\arraybackslash}p{#1}}

\journal{Journal of Information Processing and Management}

\usepackage{numcompress}

\begin{document}





\begin{frontmatter}

\title{ A Hierarchical Network-Oriented Analysis of User Participation in Misinformation Spread on WhatsApp }


\author[1]{Gabriel Peres Nobre}
\ead{gabrielnobre@dcc.ufmg.br}
\author[1,2,3]{Carlos H. G. Ferreira}
\ead{chgferreira@dcc.ufmg.br}
\author[1]{Jussara M. Almeida }
\ead{jussara@dcc.ufmg.br}

\address[1]{Departament of Computer Science, Universidade Federal de Minas Gerais, Brazil}
\address[2]{Departament of Computing and Systems, Universidade Federal de Ouro Preto, Brazil}
\address[3]{Department of Electronics and Telecommunications, Politecnico di Torino, Italy}


%
%

\begin{abstract}
WhatsApp emerged as a major communication platform in many countries in the recent years.
Despite offering only one-to-one and {\it small} group conversations, 
WhatsApp has been shown to enable the formation of a rich underlying network, crossing the boundaries of existing groups, and  with structural  properties  that favor information dissemination at large. Indeed, WhatsApp has reportedly  been used  as a forum of misinformation campaigns  with significant  social, political and economic consequences in several countries.

In this article, we aim at complementing recent studies on misinformation spread on WhatsApp, mostly focused on content properties and propagation dynamics, by looking into the {\it network} that connects users sharing the same piece of content. Specifically, we present a hierarchical network-oriented characterization of the users engaged in misinformation spread by focusing on three perspectives:  individuals, WhatsApp groups and user communities, i.e., groupings of users who, intentionally or not,  share the same content  disproportionately often. By analyzing sharing and network topological properties, our study offers valuable insights into how WhatsApp users leverage the underlying network connecting different groups to gain large reach in the spread of misinformation in the platform.


\end{abstract}

\begin{keyword}
misinformation; WhatsApp; user behavior and network analysis.
\end{keyword}

\end{frontmatter}


\section{Introduction}
During the past years, 
social media platforms have acquired a key role on the global society as a vehicle for mass information spread \cite{kumar:2018, resende:2019:WWW}. Such tools often offer a rich set of features, from one-to-one messaging to rich media content exchange \cite{jan:2011}, which ultimately contributed to engage billions of worldwide users in online activities \cite{Ferreira:2020}. It is thus of great importance to study user behavior on such platforms to be able to monitor events in the real world 
\cite{lazer:2018}.

A topic that has attracted large interest in recent years is the study of misinformation dissemination, especially in social media applications 
\cite{resende:2019:WWW, shao:2018, resende:2019:WebSci}. Indeed, it has been  argued that such applications are widely used as tools for misinformation campaigns \cite{kumar:2018}, with important negative impact on the real world during election campaigns \cite{shao:2018, wilder:2019}, humanitarian crisis \cite{yoo:2016}, pandemics \cite{pennycook:2020} and other social events \cite{allcott:2019}. Thus, the study of misinformation spread is not only a concern but more of a necessity to both understanding the phenomenon and developing methods to mitigate its negative effects \cite{wilder:2019, melo:2019}.


In such context, WhatsApp emerged as an important platform for information spread, reaching more than 2 billion users worldwide in 2020\footnote{https://www.whatsapp.com/about/}. 
 WhatsApp is a messaging application that connects users in one-to-one as well as  group conversations. Though originally designed for exchanging messages, WhatsApp has been shown to enable the formation of rich underlying networks with structural properties similar to those observed in other popular online social networks \cite{resende:2019:WWW}. These networks are fostered especially by the so called WhatsApp groups, which, though limited to only 256 simultaneous members, are linked to each other by common members which may act as bridges through which information can propagate beyond group boundaries \cite{resende:2019:WWW, nobre:2020}. 
 Not surprisingly, WhatsApp has reportedly been abused to spread misinformation \cite{digital_news_report}, mainly via group conversations, which raised major 
concerns  in many countries \cite{isaac_2021}. 

The great popularity of WhatsApp motivated a number of recent studies on different aspects of its dynamics
\cite{resende:2019:WWW, resende:2019:WebSci, melo:2019, alexandre:2020, caetano:2019}. 
In common, these prior efforts focused  on the content properties and general propagation dynamics of  pieces of information (notably misinformation).
Orthogonally, we have recently shown that WhatsApp group members do  organize themselves into communities that extrapolate the boundaries of particular groups, suggesting  that such  boundaries offer little constraint to information spread and that the underlying network indeed  conveys an important means to accelerate it \cite{nobre:2020}. However, we analyzed information dissemination in general, with no particular focus on  misinformation spread nor on how different components of this network are linked to such spread.  Thus,  the role that such  network (and its components)
plays in {\it misinformation} spread remains an open issue. 


In this article, we take a step towards filling this gap by offering an analysis of misinformation dissemination in publicly accessible WhatsApp groups focusing on the {\it  users} involved in the process. Specifically, we offer a hierarchical network-oriented characterization of user engagement in  misinformation spread from three perspectives, representing three levels of aggregation: individual users,  WhatsApp groups and user communities. As in \cite{nobre:2020}, we define a {\it community} as a grouping of users  who  shared the same content disproportionately frequently, that is, more often that could  be expected by random chance.  As such, just like individual users and formally established WhatsApp groups,  these communities, which emerge naturally from the co-sharing patterns and potentially cover users from multiple groups,  are essential components for understanding the information spread on the platform. Therefore,    our study is driven by the goal of
{\it offering valuable insights into how WhatsApp users leverage the underlying network connecting different groups to gain large reach in the spread of misinformation in the platform.}


Towards that goal, we rely on a dataset containing messages shared in 155 publicly accessible WhatsApp groups during 
6 weeks encompassing the 2018 general elections in Brazil. This was a period of great social mobilization in the country, for which there is already great amount of evidence of WhatsApp being abused for misinformation campaigns. In particular,  borrowing results from  \cite{resende:2019:WWW, resende:2019:WebSci, alexandre:2020}, the presence of misinformation is detected in textual, image and audio messages by using a dataset of  ``fake facts" previously checked by multiple Brazilian fact-checking sites. We then create a media co-sharing network model by representing users as nodes and 
connecting those who shared similar content with edges. Such networks, one for each week, can be seen as representations of potential channels for information propagation  in   WhatsApp. Thus, we investigate the role of individual users, groups and communities in misinformation spread mostly from the  perspective of their participation in these networks.



Specifically, towards identifying communities of users engaged in misinformation dissemination, we focus on the {\it backbones} of the media co-sharing networks, which includes only edges connecting users who consistently and disproportionately often shared the same content \cite{nobre:2020}.
Thus, the communities we analyze, extracted from those backbones, represent groupings of users who shared similar content more often than expected by random chance, driven by intentional behavior (i.e., by orchestration), by coincidence (as side eﬀect of the general information diﬀusion process) or by a mix of both. Beyond the network backbones, we also look into the network {\it periphery}, built from the remaining nodes and edges, which do not exhibit clear evidence of strong co-sharing patterns but still may contribute as sources of misinformation in the platform.




Our main findings can be summarized as follows:

\begin{itemize}
    \item A few users, groups and communities concentrate the top sources of misinformation spread on the platform.  Indeed, the 10 users most often engaged in misinformation spread are responsible for a large fraction (8-26\%) of all misinformation shared in a week, and also introduce a great fraction of fresh content in the platform (11.5\%). Similarly, a fraction of all groups and communities often concentrate most messages with misinformation. These groups and communities tend to have more members, thus offering a  bigger potential audience for those messages. 
    \item The media co-sharing network has an important role in misinformation spread. At individual,  group and community levels, those more often engaged in misinformation spread tend to  occupy more central positions in the network, which ultimately may favor the scale of spread. 
    \item The network backbones often  include  top misinformation sources in the network, which establish a large number of strong co-sharing connections among themselves.  On the other hand, the network periphery, which includes the largest part of the network, also includes a large fraction of users engaged in misinformation, though to a lesser degree.   
    \item The individual users mostly engaged in misinformation spread change drastically over time. Yet, they are frequently members of the  same groups. 
\end{itemize} 

The remainder of this article is organized as follows. Section \ref{related} reviews  related work while Section \ref{methodology} describes the main steps of our  methodology of analysis. Section \ref{results} presents our characterization results,  followed by our main conclusions and possible directions for future work  in Section \ref{conclusion}.

\section{Related Work}\label{related}
The analysis and modeling of information dissemination in online social networks have been the target of a large body of work, focused on the identification of influential users \cite{garadi:2018}, modeling of information cascades \cite{caetano:2019,  cheng:2016}, as well as extraction of relevant network substructures such as the network backbone \cite{Ferreira:2020, disparity_filter} and communities \cite{asim:2019, li:2015}. Most prior studies focused on traditional online social networks, tackling, for example, the dissemination of hashtags on Twitter \cite{maity:2016, rizoiu:2018} and the spread of rumors and memes on Facebook \cite{zollo:2018}. 




However, the study of misinformation spread is still a challenge \cite{kumar:2018, lazer:2018}, especially when dealing with detection, characterization and mitigation techniques for multimedia online content. The identification of misinformation content is itself a rather complex part of this challenge as it involves efforts in content analysis and truthiness detection \cite{wang:2018}. In this context, recent studies focused on assessing information credibility, characterizing the speed and strategies for misinformation spread  and the use of various features for automatic content classification \cite{vosoughi:2018, reis:2019}. Ultimately, most of the efforts are driven by the goal of mitigating, slowing down or softening misinformation effects 
in the real world \cite{lazer:2018, melo:2019}. In a complementary direction, others have  investigated the design of online tools to understand the scale of misinformation spread over the years, the psychological aspects related to misinformation content sharing and  how misinformation content is  reused to  resurface older content into news again \cite{pennycook:2020, allcott:2019}.


In light of the great popularity of existing online social networks, WhatsApp emerged as a messaging app with a rather different overall communication structure. It was originally designed with a primary focus on allowing users to exchange messages, either privately (one-to-one) or in small groups (up to 256 simultaneous members), always with end-to-end encryption. Soon after its emergence, WhatsApp rapidly attracted a large number of users in many countries. Such widespread usage was, unfortunately,   followed by several reports on the platform being  abused for misinformation campaigns   with great impact on society in Brazil \cite{wpp_abuse_brazil2}, India \cite{wpp_abuse_india} and United Kingdom \cite{wpp_abuse_uk}, to name a few examples.  
These reports, which gained great notoriety, ended up  driving the attention of the academy to the importance of WhatsApp as a platform for (mis-)information spread at large \cite{resende:2019:WWW, caetano:2019, kiran:2018}.  A number of studies emerged, relying on the existence of a large number of {\it publicly accessible WhatsApp groups}. That is, though originally a private space, a group can be made publicly accessible as the group manager shares an invitation link on the Web, effectively opening such space to the public in general. By gathering a large number of such publicly available invitation links,   researchers were able to join the groups automatically and, once a member, gather data for posterior analysis. 

Some researchers developed automatic tools to expose, in an anonymized fashion, the content being shared in WhatsApp groups \cite{kiran:2018, bursztyn:2019}, whereas others analyzed properties of such content \cite{moreno:2017, Recuero:2020, vasconcelos:2020}. The work in \cite{kiran:2018} was the first to analyze WhatsApp messages in order to uncover sharing patterns in political-oriented groups, whereas Bursztyn {\it et al.} \cite{bursztyn:2019} extended the discussion by exploring group political partisanship in  large scale. Moreover, Moreno {\it et al.} \cite{moreno:2017} studied the integration of WhatsApp into an aggregating and monitoring platform to report incidents during elections and natural events. In a different direction, the authors of \cite{Recuero:2020} characterized distinct discursive strategies applied in WhatsApp messages during the Brazilian 2018 presidential election, while Vasconcelos {\it et al.} \cite{vasconcelos:2020} investigated properties of pandemic-related YouTube videos shared by WhatsApp messages.

It is notable the focus on pieces of information previously checked as fake by fact checking agencies \cite{resende:2019:WWW, resende:2019:WebSci, melo:2019, alexandre:2020, caetano:2019, reis:2020:HKS}. For example, Resende {\it et al.} offered the first look into the properties of the content being shared on publicly accessible WhatsApp groups,  with focus on content properties and propagation dynamics of {\it images} \cite{resende:2019:WWW}. They revealed the existence of a number of images containing information priorly checked as fake, with properties quite distinct from the others. In particular, these fake images spread much faster, reaching a much larger user population, compared to other content.  More recent studies followed a similar path but focused on textual \cite{resende:2019:WebSci} and audio \cite{alexandre:2020} content, highlighting differences in content properties and propagation dynamics between (priorly checked) fake content and the rest. Caetano {\it et al.} \cite{caetano:2019}, in turn, proposed an attention cascade model to represent how  attention propagates among members of a group. They analyzed the structural and temporal properties of  cascades associated with groups that discuss political topics and with misinformation spread. They found that cascades with misinformation tend to be deeper, reach more users, and last longer in political groups than in non-political groups. Following a complementary direction, Reis {\it et al.} observed that a large fraction of the dissemination of misinformation occurred after the content was already debunked. Building on such observation, they suggested that flagging fake content could reduce the overall volume of  misinformation in the platform \cite{reis:2020:HKS}.
 
Focused mostly on the overall effects of  misinformation spread, these prior studies paid little attention to the underlying networks that connect users across different WhatsApp groups. Indeed, we are aware of only three prior efforts to shed some light into the importance of these networks to information spread \cite{resende:2019:WWW, melo:2019, nobre:2020}. 
In \cite{resende:2019:WWW},  the authors analyzed the structural properties of the network built from connecting users belonging to the same group, finding that this network has several properties, often observed in other online social networks, that have been shown to favor content virality.
Orthogonally, Melo {\it et al.} constructed the network of groups, connecting groups that have members in common,  to analyze whether limiting message forwarding could mitigate misinformation spread on WhatsApp \cite{melo:2019}. They found that, though effective in slowing down the process, such approach would not  stop  misinformation from being widely distributed.
More recently, we focused on  the media co-sharing network,  built by connecting users who shared the same content, revealing the presence of strongly connected user communities that consistently help speeding up information spread. We found that users with higher centrality in these communities are often those who contribute the most for the community’s {\it coverage} in terms of content diversity, users and groups \cite{nobre:2020}.
 
Despite shedding light into the importance of investigating the underlying networks connecting WhatsApp users and groups, these prior studies either focused on information dissemination in general, without analyzing the role of this network in the spread of misinformation  \cite{resende:2019:WWW, nobre:2020},  or studied a different, higher-level network, without focusing on the importance of its components for misinformation spread \cite{melo:2019}. In contrast, we here focus on the media co-sharing network, as it captures potential channels for information propagation,  and analyze the misinformation spread from multiple perspectives, both in terms of network component and user aggregation level.

Telegram is yet another messaging app offering individual and group communication (as well as channel broadcast) features to its users   that has gained great popularity in a number of countries\footnote{https://www.bbc.com/news/technology-55634139}, also driving the attention of academic studies.
 Notably, in \cite{nobari:2021} the authors offered a comparison of  the features available on Telegram with respect to  those of WhatsApp and Twitter, and investigated the characteristics of viral messages on the platform. In \cite{hashemi:2019}, the authors analyzed the behavior of Iranian users on Telegram by characterizing and categorizing Telegram groups in terms of their quality, whereas group quality was estimated by a number of different features. The authors of both \cite{urman:2020} and \cite{nobari:2017} offered  network-driven analyses of Telegram by building a   {\it mention network}, by connecting users who mentioned each other in their messages\footnote{On Telegram,   users and channels  have 
unique usernames which can be used as  reference to be mentioned  in a message.}. In \cite{urman:2020}, the researchers aimed at investigating  far-right networks, their actors and the properties of the communities they belong to. In contrast, in   \cite{nobari:2017}, the authors analyzed the edges of the mention network aiming at identifying  spam messages. 
 
 Our work is completely orthogonal to those prior studies on Telegram, not only because we focus on a different platform -- WhatsApp -- but also because we take a novel perspective, focusing on the role of the media co-sharing network has on user participation in misinformation dissemination.  Compared to \cite{urman:2020, nobari:2017}, which also looked at a user network,  our present analysis focuses on a different network and considers different components of this network -- backbone and periphery -- as they capture potentially different patterns of user interactions.

\begin{table}[ttt]
\centering
\fontsize{4}{4}\selectfont
\renewcommand{\tabcolsep}{1.2pt}
\begin{tabular}{|P{0.05\textwidth}|c|c|P{0.08\textwidth}|c|c|P{0.07\textwidth}|c|c|P{0.11\textwidth}|}
\hline
\textbf{Ref.} & \textbf{Year} & \textbf{System} & \textbf{Content} & \textbf{Misinformation} & \textbf{Temporal} &\multicolumn{3}{|c|}{\textbf{Network analysis}}  & \textbf{Perspective}  \\\cline{7-9}
& & & \textbf{media type} & \textbf{spread} & \textbf{analysis} &  \textbf{Global} \textbf{   view} &\textbf{Backbone} & \textbf{Periphery} & \\\hline


\cite{moreno:2017} & 2017 & WhatsApp & Text  &   &   & &  &   & Message \\ \hline

\cite{kiran:2018} & 2018 & WhatsApp & Text  &   &   & &  &   & Message \\ \hline

\cite{bursztyn:2019} & 2019 & WhatsApp & Text  &  &  & &  &  & User \\ \hline

\cite{vasconcelos:2020} & 2020 & WhatsApp & Text &  & \cmark  &  & &  & Message \\ \hline

\cite{nobre:2020} & 2020 & WhatsApp & Image, text, audio & & \cmark  &  \cmark  & \cmark  &    & Community \\  \hline 

\cite{caetano:2019} & 2019 & WhatsApp & Text & \cmark  & \cmark  & & &   & Message \\ \hline

\cite{resende:2019:WWW} & 2019 & WhatsApp & Image & \cmark  &  \cmark &  \cmark &  &   & Message \\ \hline

\cite{resende:2019:WebSci} & 2019 & WhatsApp & Text & \cmark  &  \cmark &   &  &  & Message \\ \hline

\cite{Recuero:2020} & 2020 & WhatsApp & Text & \cmark  &  &   &  & & Message \\ \hline

\cite{alexandre:2020} & 2020 & WhatsApp & Audio & \cmark  & \cmark  &   &  &  & Message \\ \hline

\cite{reis:2020:HKS} & 2020 & WhatsApp & Image & \cmark  & \cmark  &   &  &  & Message \\ \hline

\cite{melo:2019} & 2020 & WhatsApp & Image & \cmark  & \cmark  & \cmark &  &   & User, group \\ \hline


\cite{nobari:2021} & 2020 & Telegram & Text &   & \cmark  &   &  & & Message  \\ \hline

\cite{hashemi:2019} & 2019 & Telegram & Text &   & \cmark  &   &  & & Group  \\ \hline

\cite{urman:2020} & 2020 & Telegram & Text  &   & \cmark  &  \cmark &  &  & User, \\ 
& & & & & & & & & community \\ \hline


\cite{nobari:2017} & 2017 & Telegram & Text &   &   &   \cmark &  &  & Message, user \\ \hline\hline

\textbf{This work} & \textbf{2021 }& \textbf{ WhatsApp }& \textbf{Image, text, audio} & \textbf{\cmark}  & \textbf{\cmark } &  \textbf{\cmark } & \textbf{\cmark } &\textbf{\cmark}  & \textbf{User, group, community} \\ \hline
\end{tabular}


\caption{Overview of existing literature.}
\label{tab:related_wpp}
\end{table}

 We summarize how our work distinguishes from the existing literature, notably prior studies on WhatsApp and Telegram, in Table \ref{tab:related_wpp}. In this table, each row represents a prior study, and  columns 3 to 10 list different aspects defining the scope of analysis. We start by noting that  the literature is quite recent but already includes a reasonably large number of studies. Yet,  we are the first ones to analyze misinformation dissemination  by considering multimedia (text, image and audio) content\footnote{We note that, as will be described in Section \ref{subsec:dataset}, we did consider video content identifiers to build the media co-sharing networks used in our study to represent content sharing patterns. However, we did not analyze misinformation in video messages, focusing rather on misinformation in textual, image and audio content. Thus, we choose to leave {\it videos} out of  the description of our work in Table \ref{tab:related_wpp} ($4^{th}$ column).} and  network perspective, looking at  the role of both network backbone and network periphery, considering  temporal aspects, and analyzing user behavior at three different levels of aggregation. As such, we are able to uncover novel insights into user participation in misinformation spread on the platform.   

\section{ Overall Methodology}\label{methodology}

\begin{figure}[ttt!]
\centering
\begin{minipage}[c]{0.65\linewidth}
\includegraphics[width=\textwidth]{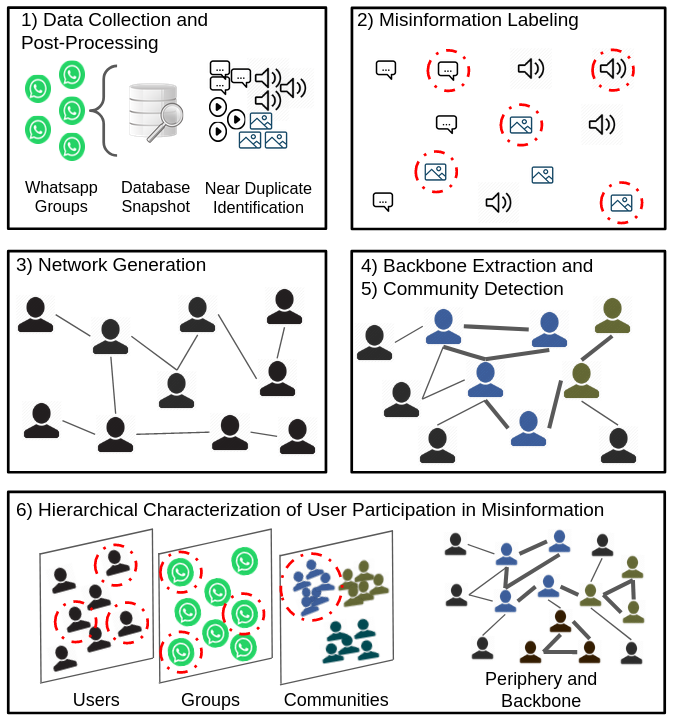}\caption{Overview of our Methodology}\label{fig:diagram}
\end{minipage}
\end{figure}


Towards characterizing user participation in misinformation spread on WhatsApp at different levels of aggregation, we adopted a methodology consisting of the steps described in Figure \ref{fig:diagram}. First, we obtained a dataset of messages shared in a number of publicly accessible WhatsApp groups. Second, we expanded the dataset by labeling a sample of messages, containing text, audio and image content, as {\it misinformation}. Third, we defined and constructed a sequence of media co-sharing networks by connecting users who shared similar content during given time intervals. We built multiple networks, enabling the study of misinformation spread during different (successive) {\it snapshots} of the platform. Then, for each network, we identified important components to the study of information dissemination, notably the {\it network backbone}, {\it communities} of nodes composing the backbone, as well as the {\it network periphery}. Finally, we analyzed the participation of users, groups and communities in misinformation spread, focusing on both sharing patterns and topological characteristics of these entities with respect to the defined networks and their main components. We further elaborate on how we performed each of these steps next\footnote{Recognizing the importance of reproducibility for scientific research, we make our (anonymized) data and the scripts for generating the networks as well as extracting backbone and communities publicly available at \url{https://zenodo.org/record/4774375\#.YNCMKfYaWV5}.}.

\subsection{Data Collection and Post-Processing} \label{subsec:dataset}



The dataset used  refers to messages shared in publicly accessible political-oriented groups in Brazil,  originally gathered by the WhatsApp Monitor \cite{resende:2019:WWW}\footnote{We thank the authors of \cite{resende:2019:WWW} for sharing the anonymized data with us.}. We restricted the  period of  analysis to six  weeks of great political mobilization in the country,  from  September \nth{17} to October \nth{28}, 2018, which includes the \nth{1} and \nth{2} rounds (October \nth{7} and \nth{28}, respectively) of the 2018 Brazilian  general elections. Our choice of the period is motivated by prior observations in academic studies \cite{resende:2019:WWW, resende:2019:WebSci, melo:2019, alexandre:2020} as well as respectable news sources\footnote{\url{https://www.washingtonpost.com/news/theworldpost/wp/2018/11/01/whatsapp-2}, \url{https://www.bbc.com/news/technology-45956557}, \url{https://www.theguardian.com/world/2019/oct/30/whatsapp-fake-news-brazil-election-favoured-jair-bolsonaro-analysis-suggests}} that the WhatsApp platform was widely used for misinformation spread in the country around this electoral period. We chose to study information spread on the monitored groups on a per-week basis, thus building six  non-overlapping snapshots. Also,  we  focused our study on the messages shared in 155 monitored groups with some message sharing activity in all six snapshots.


Table \ref{tab:wpp_data} presents a summary of the dataset for each week/snapshot, having the weeks with election dates ($3^{rd}$ and $6^{th}$) highlighted in bold. The table shows the number of unique active users, the average numbers of users and messages per group and  the average number of  messages shared by a user in a  group. The number of active users refers to number of distinct anonymized user identifiers that shared at least one message in one of the groups during the given week. The table also presents  the  numbers of messages  shared in different media types  per week. 
Despite some fluctuations in the numbers, there seems to be a general trend towards an increase in messaging activity around the election dates.

\begin{table}[ttt!]
\centering
\footnotesize\setlength{\tabcolsep}{4.5pt}
\caption{Dataset overview (155  groups, 09/17 - 10/28/2018, weeks with elections in bold).}
\label{tab:wpp_data}
\begin{tabular}{|l|c|c|c|c|c|c|}
\cline{1-7}
 {} & \multicolumn{6}{|c|}{\textbf{Weeks}} \\ 
\cline{2-7} 
\multicolumn{1}{|l|}{\textbf{Metrics  }} & 1 & 2 & \textbf{3} & 4 & 5 & \textbf{6} \\
\cline{1-7} 
\# Unique active users & 5,376 & 5,149 & 5,519 & 5,157 & 4,828 & 5,284 \\

Average \# users/group & 34.68 & 33.21 & 35.60 & 33.27 & 31.14 & 34.09 \\
Average \# messages/group & 575 & 598 & 590 & 536 & 490 & 599 \\
Average \# messsages/user (in a group) & 16.59 & 18.01 & 16.58 & 16.13 & 15.75 & 17.59 \\
\# Textual messages   & 89,136 & 92,650 & 91,438 & 83,118 & 75,982 & 92,840 \\
\# Image messages  & 13,018 & 13,208 & 13,274 & 13,471 & 11,922 & 17,113 \\
\# Audio messages   & 1,614 & 1,644 & 2,000 & 1,842 & 1,621 & 2,059 \\
\# Video messages & 10,168 & 9,515 & 9,142 & 9,508 & 9,193 & 12,344 \\
\cline{1-7}
\end{tabular}
\vspace{-0.2cm}
\end{table}

We processed the dataset to  extract and store the following
information associated with each message: timestamp, anonymized user identifier\footnote{We note that the original dataset contains only cellular phone numbers, which were  mapped onto anonymized identifiers. As such,  the same individual participating in one or more groups with multiple phone numbers will be interpreted as different users.}, 
group identifier and the  (textual, image, audio or video) message content. 
Next, we adopted a series of heuristics for filtering and identifying near duplicate content, which is a necessary step to build the media co-sharing networks, as described in Section \ref{subsec:network}. The specific heuristic employed depends on the media type. 

For textual messages, we strictly followed the procedure adopted in \cite{resende:2019:WebSci}. First, we filtered out messages shorter than 180 characters so as to retain only those with higher chance of carrying relevant information.  Towards identifying near duplicate content, the  similarity between pairs of messages was computed using the Jaccard similarity, defined as the ratio of the number of common words to the total number of distinct words in both messages. As specified in \cite{resende:2019:WebSci}, a  similarity threshold of 0.7 was empirically set by manually inspecting a sample of the messages. Thus, messages with similarity greater than 0.7 were grouped together and considered as near duplicates.  
For images, we followed the procedure specified in \cite{resende:2019:WWW}:  the  Perceptual Hashing (pHash)  algorithm \cite{phash} was used to calculate a fingerprint for each image in the dataset. This hashing algorithm is able to identify near duplicate images that hold similar color variations or that have been cropped and resized.  Images having the same hash-values were then grouped  as near duplicates. 
For audio and videos, we  used the generic identification that the WhatsApp tool creates during data transfer to detect the same content being shared multiple times. 
At the end, one message for each near duplicate group was selected as representative of the content.

We note that our near duplicate identification process is limited by the heuristics used, which, in some cases, notably audio and video content, may be somewhat conservative.  Thus, the characteristics of the networks built from our data may be more conservative than in reality.  Yet, as we will see in Section \ref{results}, even under such conservative constraints, it is still possible to identify distinguishing properties of users most often engaged in misinformation spread.  
As future work, we intend to explore more sophisticated heuristics,  such as word \cite{mikolov:2013} and sentence embeddings \cite{reimers-gurevych:2019} (for textual messages), product quantization \cite{jegou:2011} (for images) and techniques based on audio and video content analysis \cite{kordopatis-zilos:2019}.


\subsection{Misinformation Labeling} \label{sec:methodology-misinformation}

The second step of our methodology consists of identifying and labeling  messages in our dataset as misinformation. We here consider as misinformation any piece of factual content that has been priorly checked and found to be fake by a specialist, notably a fact checking agency.  As such,  messages labeled as misinformation do  contain proved fake content. However, we cannot claim anything about the truthfulness of the {\it other messages},  since they have not been checked.  Thus, we refer to them as {\it unchecked} content. 


In order to obtain a subset of messages labeled as misinformation, 
we relied on previous analyses of the dataset which had already uncovered the presence of messages containing  misinformation \cite{resende:2019:WWW, resende:2019:WebSci, alexandre:2020}.  Since these prior studies  focused on textual, image and audio content, we restricted the  messages labeled as misinformation to  these  media types.   The  same overall strategy to identify messages with misinformation was employed in those studies. First, facts previously checked as fake by at least one of  six  popular fact checking websites in Brazil were   collected. Then, the collected facts were compared with each message in our dataset. Messages considered similar enough (according to manually established thresholds)  to a previously checked fake fact were flagged as fake. For completeness, we present a brief description of the strategies adopted to identify misinformation in each media type in  \ref{appendix}, and we refer the reader to the original studies \cite{resende:2019:WWW, resende:2019:WebSci, alexandre:2020} for more details.



\subsection{Network Modeling}
\label{subsec:network}


The next step consists of building the media co-sharing networks to represent the  content sharing patterns in our WhatsApp data. A media co-sharing network is a weighted undirected graph where nodes represent users and an edge links two nodes with weight equal to the number of messages in common (i.e., near duplicates) shared by the corresponding users. We build one network for each week of data, representing a snapshot of the co-sharing patterns during that period. Network generation is performed independently of the misinformation labeling:  all (textual, image, audio and video) messages shared in each week are considered to build the corresponding network.

As defined, the media co-sharing networks expose {\it potential} channels for information spread on the platform.   The available data does not allow us to track exactly how the information propagated among users and crossed group boundaries. Also,  members of different groups may  independently learn a particular piece of information at  various sources external to WhatsApp (e.g., websites, other social networks) and chooe to share it with their groups \cite{resende:2019:WWW}.  Thus, we do not claim to faithfully represent information propagation. Rather, our network model aims at primarily representing co-sharing patterns and, through these patterns, exposing a {\it possible} means  through which specific pieces of information gain reach and visibility in the platform.


Specifically, we build a set of graphs $\mathcal{G}=\{G^{1},G^2,\ldots, G^{\Delta T}\}$, in which a given graph $G^{w}$ represents the user co-sharing patterns for week $w$ (in our case, $\Delta T = 6$). Each graph $G^w(V,E)$ is structured as follows. Each
vertex in $V$ refers to a user who posted a message during week $w$ in one of the monitored
groups. An undirected edge $e$=($v_i$, $v_j$) exists in $E$ if users corresponding to
 $v_i$ and $v_j$ shared at least one message in common during week $w$. The
weight of $e$ corresponds to the number of messages both users shared in common
during $w$.

\subsection{Network Backbone and Network Periphery Extraction}
\label{sec:method_backbone}

 The co-sharing network models {\it many-to-many} interactions that occur among multiple (possibly more than two) users at once - in our case, sharing the same media content. This kind of interaction occurs in various other domains \cite{Ferreira:2020, Benson2:2018, Newman:2018} and raises different modeling challenges if compared to often studied pairwise interactions (e.g., friendship links) \cite{Benson2:2018, Coscia:2019}.
 
In particular, as argued  in \cite{Ferreira:2020, nobre:2020, Coscia:2019}, modeling sequences of many-to-many interactions into a network
may lead to the emergence of a (potentially large) number of spurious
edges, reflecting random or sporadic user activities.
For example, consider two different scenarios:  (1) one particular viral content is massively disseminated through the WhatsApp network  as many users share it in different groups, and (2) a smaller set of users repeatedly spread similar content, reaching different audience which ultimately leads
to a large spread. We here make a distinction between the  groups of users in scenarios (1) and (2): while users in (2) exhibit strong and consistent 
co-sharing behavior, scenario (1) might reflect a large number of sporadic and random user connections.  

As in \cite{nobre:2020}, we here are particularly interested in strong and consistent co-sharing behavior, as in scenario (2), as this might reflect some intention of structural organization among users to disseminate content at large. Thus, following \cite{nobre:2020}, we  define as {\it salient} an edge connecting two users who shared the same content with a {\it disproportionately high frequency}, if compared to their co-sharing patterns with other peers. The  {\it network backbone} is then defined as the set of salient edges in the network. A  challenge  to be addressed is to identify the salient edges of a network so as to extract its backbone. 

There is a rich literature on methods to extract the backbone from
a network \cite{disparity_filter, Newman:2018, Coscia:2017}. These methods differ in terms of how salient edges are chosen (e.g., globally or on a per-node basis).  As in \cite{nobre:2020}, we employed the Disparity Filter Method \cite{disparity_filter} to extract the backbone of each media co-sharing network, as it works consistently with the aforementioned  definition of salient edge. 
Specifically,  each edge attached to a node $v_i$ is tested against the null hypothesis that
the weights of all edges incident to $v_i$ are uniformly distributed. Salient edges are those
whose weights deviate significantly from this hypothesis (thus a disproportionately high weight). Each edge is
tested twice, once for each vertex it is incident to, and it is considered salient if
it is statistically significant for both vertices when compared to a p-value (0.1 in our experiments\footnote{This p-value, which is consistent with prior studies on backbone extraction \cite{Coscia:2017},  was selected empirically by choosing, among various options, the one that led to the best tradeoff between statistical significance, number of nodes in the backbone and backbone connectivity.}).

We note that, as defined, nodes with no salient edge are {\it not} included in the backbone. Yet, these nodes may still have an important role in the spread of (mis-)information.  Thus, for the sake of analysis, we refer to all nodes that are not in the backbone as the {\it network periphery}. Whereas the backbone emphasizes collective/community behavior driven by explicit intention (e.g., orchestration) or not,  the network periphery consists   of  users acting mostly independently  or for whom there is no clear evidence of a structural organization in the network. 

\subsection{Community Detection}

As mentioned, one of our present goals is to characterize communities of users more often engaged in misinformation spread. In order to extract and characterize such communities, it is necessary to select some measurement of connectivity or similarity that captures such intuition \cite{Leskovec:2010}. We here adopt the definition of  communities as {\it non-overlapping groups} of users who share similar content with a disproportionately high frequency\footnote{Alternative definitions of communities allow for overlapping groups \cite{Guidi:2019} and dynamic structures \cite{rossetti:2018}, but these are outside of our present scope.}.
From the network perspective, each community is thus encoded as a set of nodes taken from the network backbone which is, by construction, a selection of the edges that represent  disproportionately frequent co-sharing activities.

There are several methods to detect communities, given our target definition, in the literature \cite{Fortunato:2016}. Here, we choose to identify communities in the backbone extracted from each graph $G^w$ by employing the Louvain algorithm \cite{blondel:2008}, which provides good performance in tasks with objectives similar to ours and has been widely used in different domains ranging from biological networks \cite{Rubinov:2010, Han:2016} to social media applications \cite{Ferreira:2020, nobre:2020, Ferreira:2021}. This algorithm is based on a greedy optimization of the modularity, which is a metric of quality of the communities. The modularity $Q$ is defined as: 

\begin{equation} \label{eq:modularity}
Q={\frac {1}{2m}}\sum \limits _{ij}{\bigg [}A_{ij}-{\frac {k_{i}k_{j}}{2m}}{\bigg ]}\delta (c_{i},c_{j}),
\end{equation}

\noindent where $A_{ij}$ is the weight of edge ($v_i$,$v_j$);  $k_i$ ($k_j$) is the sum of the weights of the edges attached to $v_i$ ($v_j$);  $m$ is the  sum of all of the edge weights in the graph;  $c_i$ ($c_j$) is the community assigned to $v_i$ ($v_j$); and   $\delta(c_i,c_j)=1$ if $c_i$=$c_j$ or  $0$ otherwise. The algorithm works in iterations where the nodes are assigned arbitrary communities and the resulting network partition is evaluated by the modularity metric. In the end, each node is assigned to one community and the set of communities represents an optimal modularity value. The intuition for using the Louvain algorithm is that since it maximizes the number of edges inside a community and minimizes the edges between distinct ones, users with a high media sharing similarity would end up in the same community.

\subsection{Hierarchical Characterization of User Participation in Misinformation }

We characterize user participation in misinformation spread considering three levels of aggregation: individual, group and community. The individual level refers to the actual WhatsApp users who shared at least one message in one of the monitored WhatsApp groups during the period under analysis.  The group level refers to the   WhatsApp groups that were monitored. Since the same user may join multiple WhatsApp groups simultaneously, by looking at different groups separately, we intend to investigate whether particular spaces of conversations in WhatsApp (i.e., groups) are more engaged in misinformation spread than others. The community level refers to the groupings of users with strong and consistent co-sharing patterns, extracted from the network backbones. By looking at communities, we aim at investigating whether there is evidence of  users who, potentially members of different groups, consistently share the same piece of misinformation, possibly as a strategy to gain scale of spread.

At each level of aggregation, we analyze user participation in misinformation spread from two perspectives: (1) the activity level, expressed as the total number of messages with  misinformation shared; and (2) topological properties related their connectivity in the co-sharing network. For the latter, we particularly analyze the user centrality in the network, expressed in terms of degree and closeness, and the user participation in the network backbone or in the network periphery. The closeness centrality of a node measures its average farness (inverse distance) to all other nodes \cite{Barabasi:2016}. As such, nodes with higher closeness  are able to spread information more efficiently through the network.  Our goal with such analyses is to shed some light on the role of the co-sharing network in misinformation spread, notably how users leverage the network to reach large audiences. For both group and community levels, our characterization considers the aggregation (average) of all users who are members of the group/community.

\section{Characterization Results}\label{results}

We discuss the main results from our investigation by grouping them into seven 
sections. We start by showing some statistics on the presence of misinformation in our dataset and by offering a topological description of the co-sharing networks built from the data. Next, we delve into the characterization of user participation in misinformation spread at the individual and group levels, take a look at how these elements (users/groups) are distributed across the  backbone and  the periphery of the networks, and then focus on the communities that build up the network backbones. We finish   by briefly discussing how user participation in misinformation spread evolves over time.


\subsection{Presence of Misinformation in Our Dataset}



Table \ref{tab:misinf_numbers}  shows several statistics regarding the sharing of  misinformation (identified as described in Section \ref{sec:methodology-misinformation}) during each week. 
It presents the total number of messages with misinformation and the total number of {\it distinct} messages with misinformation (content diversity). Stratified numbers for each media type are also shown. Clearly, most messages with misinformation contain textual content, although a good fraction of them are also present in images, especially in the first weeks (as already argued in \cite{resende:2019:WWW}).  Moreover, for all media types, we observe that the same misinformation content is often reshared many times  during a single week (see, for example, audios in week 2). Also, despite some fluctuations, we can see a slight trend towards an increase in volume and diversity of misinformation just before  the election dates.

Although one may consider the volume of misinformation somewhat small,  if compared to the total numbers   shown in Table \ref{tab:wpp_data}, we emphasize that those messages were shared in many groups, as will be discussed in Section \ref{sec:results_groups}, reaching a potentially large user audience who can be affected by them. Indeed, as shown  Table \ref{tab:misinf_numbers} (row 3), this potential audience, defined  as the total number of members of all groups in which a message with misinformation was shared, includes the vast majority (81\%-88\%) of all active users in the week.


\begin{table*}[tt!]
\centering
\footnotesize\setlength{\tabcolsep}{4.5pt}
\caption{Presence of misinformation in our dataset (weeks with election dates in bold). }
\label{tab:misinf_numbers}
\begin{tabular}{|l|c|c|c|c|c|c|}
\cline{1-7}
 {} & \multicolumn{6}{|c|}{\textbf{Weeks}} \\ 
\cline{2-7} 
\multicolumn{1}{|l|}{\textbf{Metrics  }} & 1 & 2 & \textbf{3} & 4 & 5 & \textbf{6} \\
\cline{1-7} 
\multicolumn{1}{|l|}{Total \# messages} & 731 & 771 & 739 & 522 & 405 & 636 \\ 
\multicolumn{1}{|l|}{\# Unique messages } & 162 & 182 & 219 & 159 & 128 & 143 \\
\multicolumn{1}{|l|}{Potential audience (\# users) } & 4550 & 4549 & 4886 & 4334 & 3926 & 4494 \\
\hline
\multicolumn{1}{|l|}{\# Texts} & 536 & 503 & 454 & 470 & 325 & 469 \\ 
\multicolumn{1}{|l|}{\# Unique texts} & 126 & 150 & 179 & 141 & 117 & 132 \\
\multicolumn{1}{|l|}{\# Images} & 176 & 136 & 157 & 22 & 16 & 6 \\ 
\multicolumn{1}{|l|}{\# Unique images} & 31 & 24 & 28 & 10 & 4 & 2 \\
\multicolumn{1}{|l|}{\# Audios} & 19 & 132 & 128 & 30 & 64 & 161 \\ 
\multicolumn{1}{|l|}{\# Unique audios} & 5 & 8 & 12 & 8 & 7 & 9 \\
\cline{1-7}
\end{tabular}
\end{table*}



\subsection{Media Co-sharing Networks}

We now turn to the co-sharing networks that model the underlying dynamics of information spread in our data. Recall that one network was built for each week, by using the results of the identification of near duplicate content.  Table \ref{tab:network_properties} shows  topological properties of the networks, notably the numbers of nodes, edges and  connected components as well as  the average and standard deviation (in parentheses) of node degree, edge weight and node clustering coefficient. 


We  note that the number of nodes (i.e., users) in the network is smaller than the total number of active users in each week (see Table \ref{tab:wpp_data}), as users who did not share any content in common with anyone else were not included in the network.  Similarly, the number of groups with at least one member in the network (last row of the table) is  smaller than the total number of groups monitored (155), as some groups did not have any co-sharing activity during a week. 

Focusing on those who did co-share content and thus were included in the networks, we can see a fairly stable number of nodes over the weeks, but a highly dynamic environment, with the number of network edges varying drastically week after week. We also see a decrease in fragmentation in the last three weeks as the number of connected components drops greatly. Overall, we observe great diversity across the nodes, in  all topological properties,  reflecting great heterogeneity in user co-sharing patterns. Nevertheless, average node degree is quite high (from 27 to 54),  indicating that, on average, a user shares similar  content with many other users. However,  the average edge weight tends to be small (up to 1.85),  suggesting that those co-sharing connections may often represent a sporadic behavior (e.g., one message shared in common). Finally,  the  clustering coefficient values (around 0.60, on average) along with the number of connected components suggest a  highly clustered network,  hinting at the  formation of strong clusters of nodes (e.g., user communities).

\begin{table*}[tt!]
\centering
\footnotesize\setlength{\tabcolsep}{2.5pt}
\caption{Topological properties  of the co-sharing networks (CC = clustering coefficient;  $\sigma$ = standard deviation). Weeks with election dates in bold. }
\label{tab:network_properties}
\begin{tabular}{|l|c|c|c|c|c|c|}
\cline{1-7}
 {} & \multicolumn{6}{|c|}{\textbf{Weeks}} \\ 
\cline{2-7} 
\multicolumn{1}{|l|}{\textbf{Metrics}} & 1 & 2 & \textbf{3} & 4 & 5 & \textbf{6} \\
\cline{1-7} 
\multicolumn{1}{|l|}{\# Nodes} & 1,936 & 1,842 & 2,002 & 2,091 & 1,768 & 2,104 \\ 
\multicolumn{1}{|l|}{\# Edges} & 26,207 & 32,370 & 28,297 & 52,085 & 24,008 & 57,354 \\
\multicolumn{1}{|l|}{\# Conn. components } & 50 & 53 & 52 & 29 & 36 & 33  \\
\multicolumn{1}{|l|}{Avg. degree ($\sigma$)} & 27.07 (41) & 35.15 (49) & 28.27 (38) & 49.82 (69) & 27.16 (39) & 54.52 (81) \\
\multicolumn{1}{|l|}{Avg. edge weight ($\sigma$)} & 1.20 (0.96) & 1.44 (1.65) & 1.38 (1.51) & 1.59 (2.51) & 1.30 (1.28) & 1.85 (2.37) \\
\multicolumn{1}{|l|}{Avg. CC ($\sigma$)} & 0.58 (0.38) & 0.61 (0.37) & 0.60 (0.37) & 0.65 (0.34) & 0.60 (0.36) & 0.62 (0.35) \\
\multicolumn{1}{|l|}{\# Groups covered} & 142 & 137 & 141 & 136 & 133 & 133 \\ 
\cline{1-7}
\end{tabular}
\end{table*}

Having presented the statistics on misinformation and the topological properties of the  co-sharing networks that serve as background for our study, we are then ready to discuss our characterization results, starting with the analysis at the individual level. Throughout the rest of this section, aiming at maintaining consistency through all our analyses, we focus on users who have co-shared   content with someone else, that is, users  included in the co-sharing networks. 
 
\subsection{Individual Users}

Table \ref{tab:user_misinf} provides a summary of statistics regarding individual users engaged in misinformation spread. It presents the number and the percentage of users who shared at least one message with misinformation in each week\footnote{ In this table, as in the others that follow it, we use the term {\it mmsg} to refer to messages with misinformation.}. As shown, an expressive fraction (from 15\% to 27\%) of all users who shared some content during a week also shared some misinformation during that period, again with a slight increase in this  activity in the weeks around election dates.

\begin{table*}[tt!]
\centering
\footnotesize\setlength{\tabcolsep}{2pt}
\caption{Individual users engaged in misinformation spread (mmsg = messages with misinformation; $\sigma$ = standard deviation; rows 3-6 computed for users who shared misinformation).}
\label{tab:user_misinf}
\begin{tabular}{|l|c|c|c|c|c|c|}
\cline{1-7}
 {} & \multicolumn{6}{|c|}{\textbf{Weeks}} \\ 
\cline{2-7} 
\multicolumn{1}{|l|}{\textbf{Metrics}} & 1 & 2 & \textbf{3} & 4 & 5 & \textbf{6} \\
\cline{1-7} 
\multicolumn{1}{|l|}{\#  Users with $\geq$1 mmsg} & 373 & 509 & 506 & 380 & 275 & 424 \\   
\multicolumn{1}{|l|}{\% All active  users} & 19\% & 27\% & 25\% & 18\% & 15\% & 20\%  \\  
\multicolumn{1}{|l|}{Avg \# mmsgs/user ($\sigma$)  } & 1.95(3.86)  & 1.51(1.07)  & 1.46(1.07) & 1.37(0.79)  & 1.47(1.29) & 1.5(1.22) \\
\multicolumn{1}{|l|}{Max \# mmsgs/user} & 56 & 11 & 12 & 8 & 16 & 15 \\  
\multicolumn{1}{|l|}{Avg \# uniq. mmsgs/user ($\sigma$) } & 1.33(0.80)  & 1.40(0.83)   & 1.33(0.81)   & 1.29(0.60)   & 1.32(0.75)  & 1.34(0.92)   \\
\multicolumn{1}{|l|}{Max. \# unique mmsgs/user} & 8 & 7 & 7 & 5 & 6 & 11 \\
\cline{1-7}
\end{tabular}
\end{table*}

Focusing on users who shared at least one message with misinformation in each week, Table \ref{tab:user_misinf}  shows that, on average, they shared a small number of such messages (below 2) per week (row 3). Yet, the large standard deviations suggest great diversity in user behavior, with some users sharing as many as 16 (week  5) or 56 (week 1) messages with misinformation in one week (row 4). The table also shows a smaller average (and maximum) numbers of {\it unique } messages with misinformation (rows 5 and 6), indicating that some users   share the same piece of misinformation multiple times during the same week.

\begin{figure*}[ttt]
\centering
    \begin{center}
        \begin{subfigure}[t]{0.48\linewidth}
            \includegraphics[width=\linewidth]{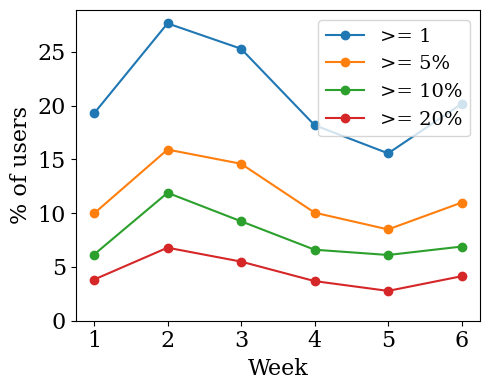}
            \caption{Level of participation in misinformation}
            \label{fig:misinf_users_relative}
        \end{subfigure}
        \begin{subfigure}[t]{0.48\linewidth}
          \includegraphics[width=\linewidth]{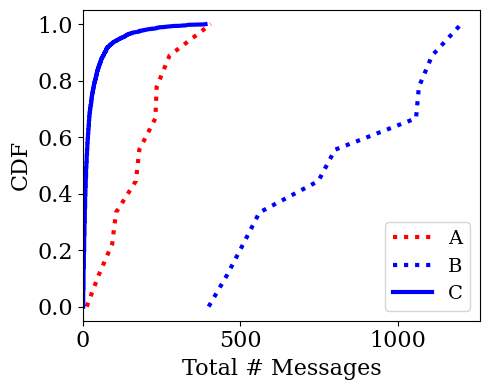}
          \caption{\# Messages/user}
            \label{fig:user_cdf_messages}
        \end{subfigure}
        
        \begin{subfigure}[t]{0.48\linewidth}
          \includegraphics[width=\linewidth]{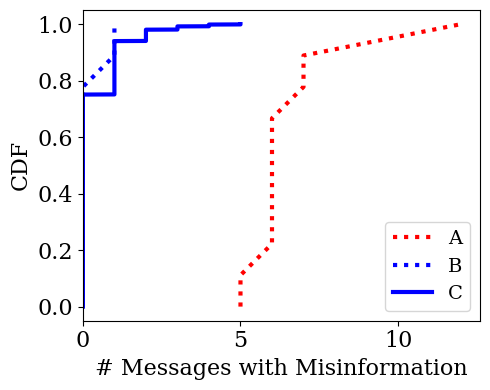}
          \caption{\# Messages with misinformation/user}
            \label{fig:user_cdf_misinformation}
        \end{subfigure}
        \begin{subfigure}[t]{0.48\linewidth}
          \includegraphics[width=\linewidth]{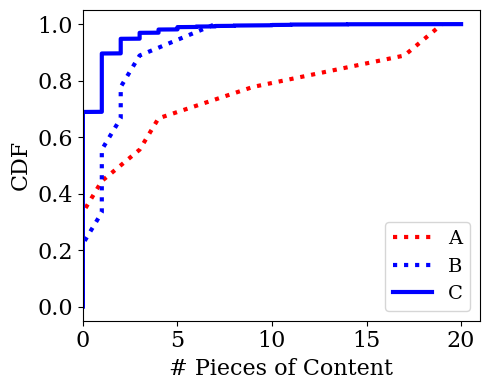}
          \caption{Introduction of new content}
            \label{fig:user_cdf_first_media}
        \end{subfigure}
        \caption{Content sharing patterns of individual users (Figures (b-d): results for week 3;   curves (A): top-10 misinformation users, (B): top-10 all, (C): others).}
        \label{fig:user_results2}
    \end{center}
\end{figure*}

We  delve deeper into the sharing patterns of individual users by analyzing the results in Figure \ref{fig:user_results2}. Figure \ref{fig:misinf_users_relative} shows the  weekly percentage of users engaged in misinformation spread for different {\it levels of participation}.  The level of participation of a user at week $w$ is defined based on the number of messages with misinformation shared by her during $w$.  The curve labeled as {\it $\geq$ 1} refers to all users who sent at least one message with misinformation (same result in Table \ref{tab:user_misinf}), whereas the curves labeled as {\it $\geq$ x\%} refers to users for whom at least x\% of all messages shared by them were misinformation (larger values of $x$ suggest greater engagement). As the figure shows, these percentages vary across the weeks, reaching local peaks around the  election dates. We also note that a non-negligible fraction of  users (up to 7\%) were  more heavily engaged in this activity, with at least 20\% of the content shared by them being misinformation.

Given such diversity in the level of user participation in misinformation spread, we chose to continue our investigation by looking at three different categories of users separately. For each week $w$,  we grouped users into:   (1) {\it top misinformation users:} top-$k$ users who shared the largest number of messages with misinformation during $w$; (2) {\it top all:} top-$k$ users who shared the largest number of messages in general  during $w$ (excluding those in (1)); and (3) {\it others:} the remaining users. 
We chose $k$=$10$ so as to include in (1) users who severely engaged in misinformation spread\footnote{We observed similar qualitative results for k=10, 20, 50 and 100.}
. Indeed, collectively, these 10 users (out of hundreds) were responsible for 8-26\% of all misinformation shared in a week. 


Figures \ref{fig:user_cdf_messages} and \ref{fig:user_cdf_misinformation} show the cumulative distribution functions (CDFs) of the  total number of messages and the number of messages with misinformation shared by a user,  for each user category. These two graphs refer to sharing activity in week 3, but the results for the other weeks are similar.  As shown in the figures, the top misinformation users (curve $A$) did not share as many messages as the top all users (curve $B$), who were highly active users exceeding 1000 messages (mostly content not priorly identified as misinformation) in a week. 
Yet, both groups of top users (curves $A$ and $B$) clearly share many more messages than the other users (curve $C$). 
The top-10 misinformation users also distinguish themselves from the other two categories  (curves $B$ and $C$) in terms of the volume of misinformation shared. Moreover as shown in Figure \ref{fig:user_cdf_first_media}, the top misinformation users were responsible for introducing new content  much more often than users in the other two sets. This implies that these users not only shared more misinformation but they were often the first ones to bring this content to the WhatsApp network (as observed in our data).

\begin{figure*}[ttt!]
\centering
    \begin{center}
        \begin{subfigure}[t]{0.32\linewidth}
          \includegraphics[width=\linewidth]{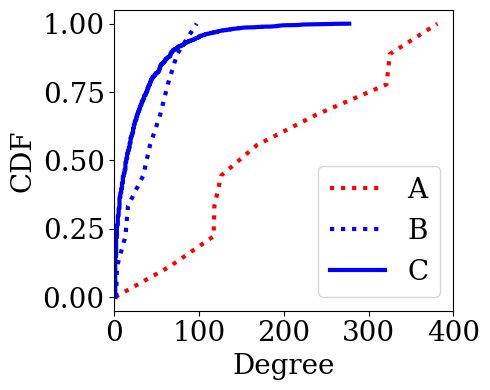}
          \caption{ Node degree}
            \label{fig:user_cdf_degree}
        \end{subfigure}
        \begin{subfigure}[t]{0.32\linewidth}
          \includegraphics[width=\linewidth]{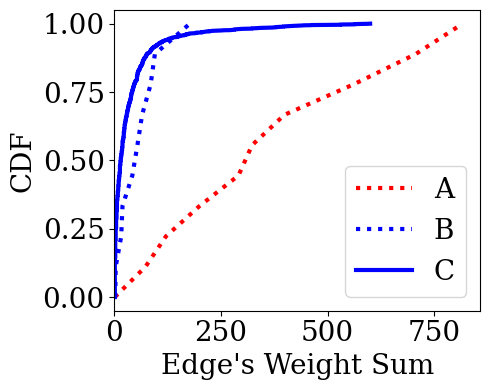}
          \caption{Total edge weight}
            \label{fig:user_cdf_weight}
        \end{subfigure}
        \begin{subfigure}[t]{0.32\linewidth}
            \includegraphics[width=\linewidth]{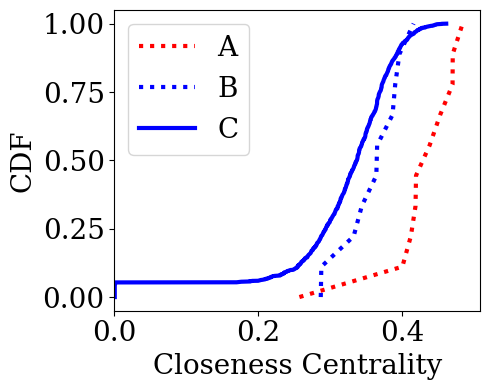}
            \caption{Closeness}
            \label{fig:user_cdf_std_closeness}
        \end{subfigure}
        \caption{Topological properties of users  in the co-sharing networks (Results for week 3; curves (A): top-10 misinformation users, (B): top-10 all, (C): others).}
        \label{fig:user_results3}
    \end{center}
\end{figure*}

We now look into user participation in misinformation spread from the perspective of the user's role in the media co-sharing network.  Figure \ref{fig:user_results3} shows the CDFs of several topological properties associated with users (nodes) in each of the three categories, top-10 misinformation users (curve A), top-10 all users (curve B) and other users (curve C),  considering once again the network built for week 3. Very similar results were obtained for the other weeks, being thus omitted.   In general, we see a clear distinction of the top misinformation users from the other two categories of users, which, in turn, exhibit somewhat similar topological patterns.

As Figures \ref{fig:user_cdf_degree} and \ref{fig:user_cdf_weight} show, the top misinformation users tend to have  much higher degree and much higher sum of edge weights in the network than users in the other two categories. That is, the {\it co-sharing} activities, when multiple users share the same piece of content, occur much more often among the top misinformation users, both in terms of number of co-sharing peers (degree) and number of messages shared in common (edge weight). 
Also, in Figure \ref{fig:user_cdf_std_closeness}, we show that these users tend to occupy much more central roles in the network topology, as expressed by their closeness centrality. In this context, it means that the top misinformation users are closer to the others in the network and, therefore, demonstrate a higher efficiency in disseminating a given piece of information across the network.
Overall, these results suggest that users more often engaged in misinformation spread do position themselves strategically in the co-sharing network: by establishing a larger number of stronger connections, they end up occupying more central positions in the network, which ultimately may favor misinformation spread. 

\subsection{Groups} \label{sec:results_groups}
We now analyze the participation of WhatsApp groups, as a collection of individuals discussing in a common (virtual) space, in misinformation spread. As done for users, we analyze group participation during each week separately. Table \ref{tab:group_misinf} shows some general statistics by presenting, for each week of analysis, the number of groups with at least one message with misinformation, the percentage of these groups with respect to all groups where co-sharing occurred during the week, as well as the average, standard deviation and the maximum number of user members and messages with misinformation in these groups.



We start by noting that some misinformation was spread in   most (up to 78\%) groups in all monitored weeks (row 2). Indeed, the number of messages with misinformation shared weekly on each such group varied from 5 to 8, on average (row 3). Yet, the sharing of misinformation was much more intense in some of these groups, reaching up to 202 messages in a group during a single week (row 4). Table \ref{tab:group_misinf} also shows that the  potential audience of such messages can be expressive. On average, the number of user members per group remains roughly stable in all weeks, falling around 20 users per group (row 5), although some groups may have as many as 107 members during a week (row 6).

\begin{table*}[t!]
\centering
\footnotesize\setlength{\tabcolsep}{2.5pt}
\caption{Groups engaged in misinformation spread (mmsg = message with misinformation; $\sigma$ = standard deviation; rows 3-6 computed for groups where some misinformation was shared).}
\label{tab:group_misinf}
\begin{tabular}{|l|c|c|c|c|c|c|}
\cline{1-7}
 {} & \multicolumn{6}{|c|}{\textbf{Weeks}} \\ 
\cline{2-7} 
\multicolumn{1}{|l|}{\textbf{Metrics}} & 1 & 2 & \textbf{3} & 4 & 5 & \textbf{6} \\
\cline{1-7} 
\multicolumn{1}{|l|}{\#  Groups with $\geq$1 mmsg } & 91 & 108 & 101 & 91 & 80 & 96 \\ 
\multicolumn{1}{|l|}{\%  Groups} & 64\% & 78\% & 71\% & 66\% & 60\% & 72\%  \\
\multicolumn{1}{|l|}{Avg. \# mmsgs/group($\sigma$)} & 8.03 (21) & 7.13 (8) & 7.31 (7) & 5.73 (6) & 5.06 (4) & 6.62 (6) \\
\multicolumn{1}{|l|}{Max. \# mmsgs/group} & 202 & 61 & 44 & 36 & 21 & 31 \\
\multicolumn{1}{|l|}{Avg. \# users/group ($\sigma$)} & 21.4 (19) & 17.7 (17) & 20.2 (18) & 22.6 (20) & 21.9 (17) & 21.6 (19) \\
\multicolumn{1}{|l|}{Max. \# users/group} & 107 & 92 & 93 & 102 & 73 & 92 \\
\cline{1-7}
\end{tabular}
\end{table*}

\begin{figure*}[!ttt]
\centering
    \begin{center}
        \begin{subfigure}[t]{0.48\linewidth}
          \includegraphics[width=\linewidth]{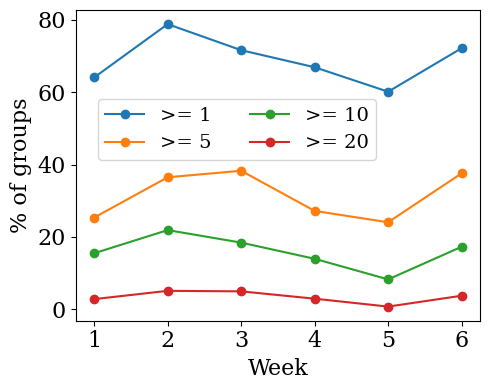}
          \caption{Level of participation in misinformation}
            \label{fig:group_misinf_relative}
        \end{subfigure}
        \begin{subfigure}[t]{0.48\linewidth}
            \includegraphics[width=\linewidth]{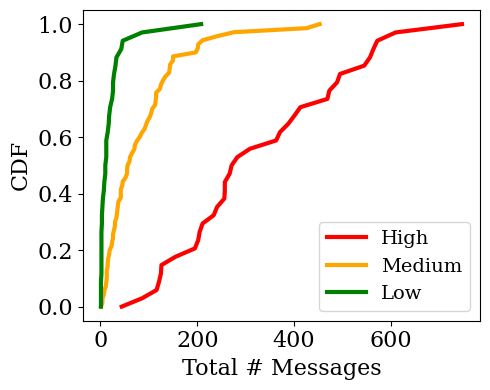}
            \caption{\# Messages/group}
            \label{fig:group_message}
        \end{subfigure}
        
        \begin{subfigure}[t]{0.48\linewidth}
          \includegraphics[width=\linewidth]{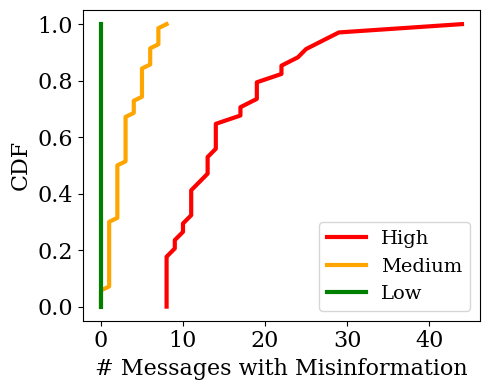}
          \caption{\# Messages with misinformation/group}
            \label{fig:group_misinf_count}
        \end{subfigure}
        \begin{subfigure}[t]{0.48\linewidth}
          \includegraphics[width=\linewidth]{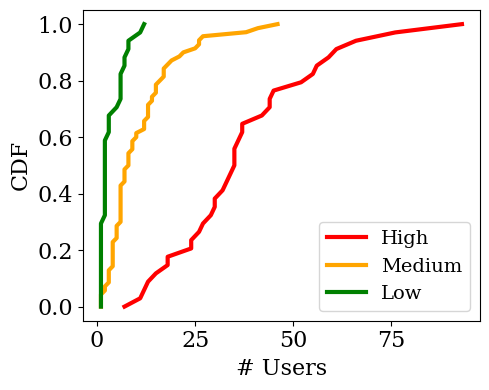}
          \caption{\# Members/group}
            \label{fig:group_user_count}
        \end{subfigure}
        \caption{Content sharing patterns of WhatsApp groups (Figures (b-e): results for week 3).}
        \label{fig:group_results1}
    \end{center}
\end{figure*}

Figure \ref{fig:group_results1} shows content sharing 
patterns of  the  WhatsApp groups. Specifically, Figure 
\ref{fig:group_misinf_relative} shows the weekly percentages of 
groups where the number of messages of misinformation shared 
during  the  week exceeds a given threshold (1, 5, 10 or 20 messages). Clearly, there is great diversity across the groups.  For 62-76\% of the 
groups, the level of misinformation sharing may be considered small (fewer than 5 messages in a week).  Yet, 
the sharing of misinformation may be considered 
large (more than 10 messages) in 8-21\% of the groups,  and quite intense in a few of them,
 with
more  20 messages shared in a single week.

Given such group diversity, once again we chose to analyze groups
separately. During each week $w$, we define three group categories: (1) {\it high}: top 25\% groups with the largest numbers of messages with misinformation shared during $w$; (2) {\it low}: bottom 25\% groups with the smallest numbers of messages with misinformation; and (3) {\it medium}: the remaining groups.

Figures \ref{fig:group_message} and  
\ref{fig:group_misinf_count}
show the distributions of the  number of messages 
and number of messages with misinformation shared
per group during a single week (week 3) for each group category. As shown, groups in the {\it high} category 
 not only share the largest volumes of misinformation (71-84\% of all misinformation shared in the week) but also tend to have the most intense sharing activity in general. Indeed, we did find a strong positive correlation between these two measures for all
 groups (Spearman correlation $\rho$=0.83).  
 Moreover, as shown in Figure 
 \ref{fig:group_user_count}, groups with higher volume of misinformation 
 tend to have a larger number of user members, thus offering a potentially larger audience to such messages. Again, we found a strong correlation between the number of messages with misinformation and the number of members in the group ($\rho$=0.79). Obviously, 
 some of these
 users are those sharing these messages. However, we note that a large fraction of  the group members, specifically 40-85\% of members of  groups in the high category,  did {\it not} share any misinformation, and yet were potential target of this type content.  Similar results were obtained for all weeks.

\begin{figure*}[!tt]
\centering
    \begin{center}
        \begin{subfigure}[t]{0.32\linewidth}
            \includegraphics[width=\linewidth]{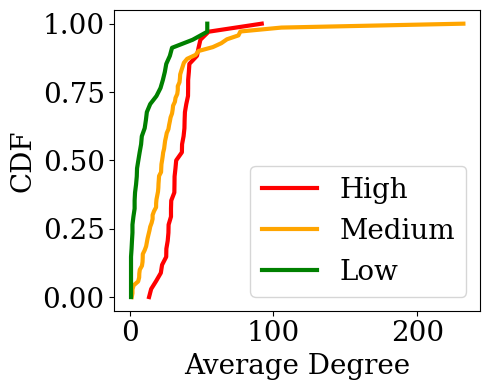}
            \caption{Node degree}
            \label{fig:group_cdf_degree}
        \end{subfigure}
        \hfill
        \begin{subfigure}[t]{0.32\linewidth}
            \includegraphics[width=\linewidth]{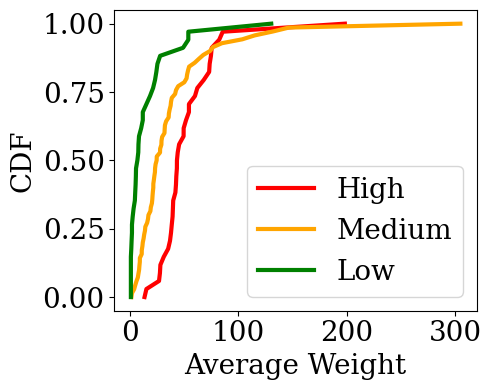}
            \caption{Total edge weight}
            \label{fig:group_cdf_edge_weight}
        \end{subfigure}
        \hfill
        \begin{subfigure}[t]{0.32\linewidth}
            \includegraphics[width=\linewidth]{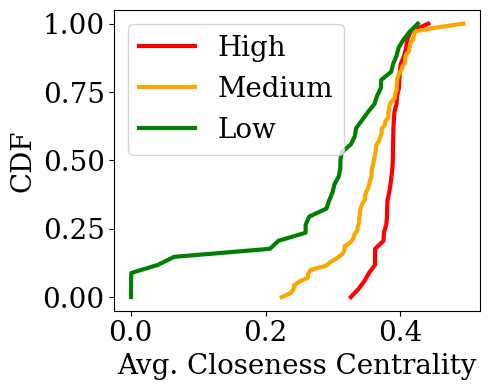}
            \caption{Closeness}
            \label{fig:group_cdf_closeness}
        \end{subfigure}
        \caption{Topological properties of groups in the co-sharing networks (Average results computed for all  group members for week 3).}
        \label{fig:group_results2}
    \end{center}
\end{figure*}

We also look into the role of groups in the co-sharing networks.  Figure \ref{fig:group_results2} shows the distributions of several topological properties for groups in each category for week 3. Again, results for the other weeks are very similar. As groups are not explicitly modeled in the networks, we compute each metric by taking the average over all group members as representative of the group. The figure shows results very consistent to the observations at the individual level (Figure \ref{fig:user_results3}). Overall, we find that groups with higher volumes of misinformation  tend to have users with larger node degrees (Figure \ref{fig:user_cdf_degree}),  (somewhat) larger total sum of weight (Figure \ref{fig:user_cdf_weight}) and more centrally located in the network (Figure \ref{fig:user_cdf_std_closeness}). These results seem to  suggest that such groups may have had a higher ability to disseminate misinformation in WhatsApp by leveraging the connectivity of their members in the network.

\subsection{Participation in the Network Backbone and in the Network Periphery}

Before analyzing user engagement in misinformation at the level of communities, we delve further into the role that individual users and groups have in the co-sharing network by analyzing their participation in the network backbone and in the network periphery. As argued in Section \ref{sec:method_backbone}, the  
backbone emerges mostly from consistent and strong
collective  behavior (co-sharing), whereas the network periphery consists of users acting mostly independently or with no clear evidence of structural organization in the network. Thus by investigating user participation in these two components of the network, we offer insights into potential behavioral patterns driving misinformation dissemination.  



We start by presenting, in Table \ref{tab:backbone_properties},  a topological description of the backbones extracted from the networks of each week. Note that the backbones are significantly smaller than the original networks, in numbers of  nodes  and edges (see network sizes in Table \ref{tab:network_properties}). Thus, many weak edges (and nodes) were indeed removed to uncover the backbones, falling in the network periphery. Indeed, as shown in Table \ref{tab:backbone_properties} (row 7), only 10-20\% of all nodes/users in the network lie in the backbone (the rest lie in the periphery). However, the backbones cover between 52\% and 71\% of all groups (last row), that is, at least one user in these groups is in the backbone. Conversely, some groups (29\%-48\%) have all their members in the network periphery.

\begin{table*}[t!]
\centering
\footnotesize\setlength{\tabcolsep}{2.4pt}
\caption{Topological properties of the network backbones (CC = clustering coefficient; $\sigma$ =
standard deviation). Weeks with election dates in bold.}
\label{tab:backbone_properties}
\begin{tabular}{|l|c|c|c|c|c|c|}
\cline{1-7}
 {} & \multicolumn{6}{|c|}{\textbf{Weeks}} \\ 
\cline{2-7} 
\multicolumn{1}{|l|}{\textbf{Metrics}} & 1 & 2 & \textbf{3} & 4 & 5 & \textbf{6} \\
\cline{1-7} 
\multicolumn{1}{|l|}{\# Nodes} & 212 & 297 & 229 & 317 & 230 & 435 \\ 
\multicolumn{1}{|l|}{\# Edges} & 556 & 1173 & 970 & 2270 & 851 & 4229 \\
\multicolumn{1}{|l|}{\# Conn. components} & 11 & 19 & 19 & 8 & 14 & 11 \\ 
\multicolumn{1}{|l|}{Avg. degree ($\sigma$)} & 5.3 (6.7) & 7.9 (10.2) & 8.5 (10.8) & 14.3 (17.9) & 7.4 (9.0) & 19.4 (21.5) \\
\multicolumn{1}{|l|}{Avg. edge weight ($\sigma$)} & 5.9 (3.8) & 7.6 (4.7) & 7.4 (3.3) & 11.1 (6.1) & 6.6 (3.6) & 8.3 (4.7) \\
\multicolumn{1}{|l|}{Avg. CC ($\sigma$)} & 0.40 (0.41) & 0.42 (0.40) & 0.44 (0.41) & 0.56 (0.38) & 0.46 (0.40) & 0.57 (0.37) \\
\hline
\multicolumn{1}{|l|}{User coverage} & 10\% & 16\% & 11\% & 15\% & 13\% & 20\% \\
\multicolumn{1}{|l|}{Group coverage} & 52\% & 57\% & 53\% & 59\% & 63\% & 71\% \\
\cline{1-7}
\end{tabular}
\end{table*}

As consequence of the removal of a large fraction  of the edges, nodes in the backbone have much smaller average degrees (row 4), but also much stronger edge weights (row 5), besides building up fewer connected components (row 3).  Moreover, as many nodes were also removed, the  average clustering coefficients (row 6) are somewhat smaller compared to the original networks, consistently with prior studies \cite{disparity_filter}.  Overall, the backbones are much smaller graphs, containing fewer connected components built from much stronger node connections.


Table \ref{tab:backbone_misinf_metrics} offers an overview of the participation of the backbone in misinformation spread by presenting statistics on the users who shared misinformation and are located in the backbones,  as well as the groups where some misinformation was shared and with members located in the backbone. We note the presence of often more than 100  users  in the backbone who shared some misinformation (row 1). In relative terms, often around half of all users in the backbone shared some misinformation (row 2). Moreover, despite representing around  only one third (23-41\%) of all users who shared some misinformation in the network (row 3), these users in the backbone are often among the top-10 and top-50 users most engaged in misinformation spread during the week (rows 4 and 5). Note that  these users in the backbone are responsible for a large fraction  (32-53\%) of all misinformation shared  (row 6), and, importantly, up to 60\% of all diversity  in misinformation  (row 7).   
Also, we note that users in the backbone who share misinformation  are members of up to 68 different groups (row 8), which represent the most groups covered by the backbone (row 9) as well as  most groups where some misinformation was shared (row 10). 

\begin{table*}[t!]
\centering
\footnotesize\setlength{\tabcolsep}{2.5pt}
\caption{Misinformation sharing by users and groups covered by the backbone.  }
\label{tab:backbone_misinf_metrics}
\begin{tabular}{|l|c|c|c|c|c|c|}
\cline{1-7}
 {} & \multicolumn{6}{|c|}{\textbf{Weeks}} \\ 
\cline{2-7} 
\multicolumn{1}{|l|}{\textbf{Metrics}} & 1 & 2 & \textbf{3} & 4 & 5 & \textbf{6} \\
\cline{1-7} 
\multicolumn{1}{|l|}{\# Users in backbone who shared misinformation} & 105 & 170 & 120 & 144 & 95 & 175 \\
\multicolumn{1}{|l|}{\% Users in backbone who shared misinformation } & 49\% & 57\% & 52\% & 45\% & 41\% & 40\% \\
\multicolumn{1}{|l|}{\% Users who shared misinformation who are in backbone} & 28\% & 33\% & 23\% & 37\% & 34\% & 41\% \\
\multicolumn{1}{|l|}{\% Top-10 misinformation users who  are in backbone} & 40\% &  100\% & 80\% & 70\% & 90\% & 100\% \\
\multicolumn{1}{|l|}{\% Top-50 misinformation users who  are in backbone} & 64\% & 90\% & 64\% & 72\% & 74\% & 82\% \\
\hline
\multicolumn{1}{|l|}{\% Messages with misinformation } & 35\% & 46\% & 32\% & 44\% & 47\% & 53\% \\
\multicolumn{1}{|l|}{\% Unique messages with misinformation} & 52\% & 52\% & 43\% & 52\% & 54\% & 60\% \\ \hline
\multicolumn{1}{|l|}{\# Groups with misinformation covered by backbone} & 50 & 61 & 54 & 59 & 54 & 68 \\
\multicolumn{1}{|l|}{\% Groups covered by backbone with misinformation} & 66\% & 77\% & 71\% & 72\% & 63\% & 71\% \\
\multicolumn{1}{|l|}{\% Groups with misinformation covered by backbone} & 54\% & 56\% & 53\% & 64\% & 67\% & 70\% \\ 
\cline{1-7}
\end{tabular}
\end{table*}

All these statistics attest for the importance of the backbone to the dissemination of misinformation. Despite covering only around 15\% of all users in the network, the backbone nodes, who include most top misinformation sources in the network, are responsible for a great part of all misinformation spread in the platform, with  a clear dominance in terms of content diversity. The strong co-sharing patterns linking these users in the backbone suggest the possibility of coordinated effort to gain audience scale. 
In contrast, we also note the large presence of  users engaged in misinformation spread, though  with a lesser degree,   in the network periphery. Indeed, representing roughly two thirds of all users who shared some misinformation in a single week, they are responsible for a large fraction  (up to 68\%) of all misinformation shared in the period. Next,  we delve further into the network backbones by characterizing the participation of communities, detected in these backbones, in misinformation spread.

\subsection{Communities}

We finally reach the third level of user aggregation and study the participation of user communities in misinformation spread. As shown in Table \ref{tab:comm_misinf}, we detected between 14 and 27  well structured (i.e., large modularity) communities over the weeks (rows 1 and 2).  Most communities (63-85\%) had some participation in misinformation sharing (rows 3 and 4). Indeed, the numbers of messages with misinformation shared by a community (row 5) largely exceed those of a group (Table \ref{tab:group_misinf}), despite the typically smaller number of members in a community (row 7), at least on average.  Yet, once again we see great diversity across communities, in terms of membership size and engagement of these members in misinformation spread. Some communities exceed 100 members (row 8). This is quite impressive, given that, by definition, these users have strong co-sharing patterns. Also, some communities shared, in a single week, as many as 122 messages with misinformation (row 6).  Note also how communities connect together users across many different groups (rows 9 and 10). 

\begin{table*}[t!]
\centering
\footnotesize\setlength{\tabcolsep}{2.5pt}
\caption{Communities engaged in misinformation spread (mmsg = message with misinformation; $\sigma$ = standard deviation;)}
\label{tab:comm_misinf}
\begin{tabular}{|l|c|c|c|c|c|c|}
\cline{1-7}
 {} & \multicolumn{6}{|c|}{\textbf{Weeks}} \\ 
\cline{2-7} 
\multicolumn{1}{|l|}{\textbf{Metrics}} & 1 & 2 & \textbf{3} & 4 & 5 & \textbf{6} \\
\cline{1-7} 
 \multicolumn{1}{|l|}{\# communities detected} & 19 & 27 & 26 & 14 & 23 & 21 \\
\multicolumn{1}{|l|}{Modularity} & 0.69 & 0.61 & 0.51 & 0.53 & 0.61 & 0.56 \\ \hline
\multicolumn{1}{|l|}{\# communities with   mmsg} & 12 & 18 & 19 & 12 & 15 & 17 \\
\multicolumn{1}{|l|}{\% communities with  mmsg} & 63\% & 66\% & 73\% & 85\% & 65\% & 80\% \\
\multicolumn{1}{|l|}{Avg \#mmsgs/comm. ($\sigma$)} & 21.4 (33) & 19.7 (27) & 12.7 (19) & 19.5 (31) & 12.8 (12) & 19.9 (30) \\
\multicolumn{1}{|l|}{Max. \#mmsgs/comm.} & 122 & 96 & 59 & 109 & 34 & 104 \\
\multicolumn{1}{|l|}{Avg \#users/comm. ($\sigma$)} & 16.4 (24) & 15.4 (19) & 11.2 (14) & 25.6 (37) & 14.0 (12) & 25.0 (34)  \\
\multicolumn{1}{|l|}{Max. \#users/comm.} & 91 & 64 & 51 & 130 & 40 & 118 \\ 
\multicolumn{1}{|l|}{Avg. \#groups/comm. ($\sigma$)} & 12.0 (12) & 12.5 (13) & 10.0 (11) & 16.4 (17) & 14.0 (11) & 18.8 (19) \\ 
\multicolumn{1}{|l|}{Max. \#groups/comm.} & 46 & 44 & 40 & 54 & 33 & 55 \\ 
\cline{1-7}
\end{tabular}
\end{table*}




As before, we also categorize communities for further analysis.  For each week $w$, we define three  categories: (1) {\it high}: consisting of the top 25\% communities in volume of misinformation shared; (2) {\it low}: consisting of the bottom 25\%; and (3) {\it medium}: the remaining 50\% of them.  
 
 \begin{figure*}[!t]
\centering
    \begin{center}
        \begin{subfigure}[t]{0.48\linewidth}
          \includegraphics[width=\linewidth]{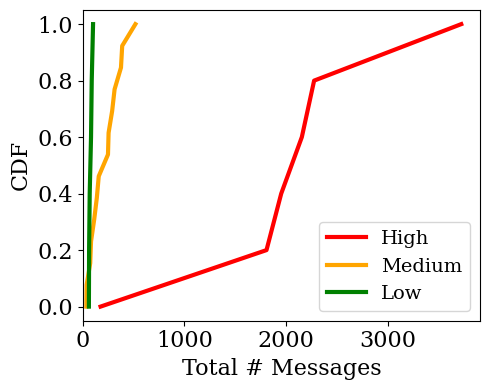}
          \caption{\# Messages/community}
            \label{fig:comm_messages}
        \end{subfigure}
        \begin{subfigure}[t]{0.48\linewidth}
          \includegraphics[width=\linewidth]{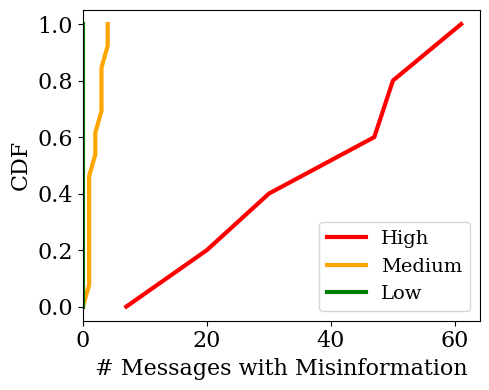}
          \caption{\# Messages w. misinformation/community}
            \label{fig:comm_misinf}
        \end{subfigure}
        
        \begin{subfigure}[t]{0.48\linewidth}
          \includegraphics[width=\linewidth]{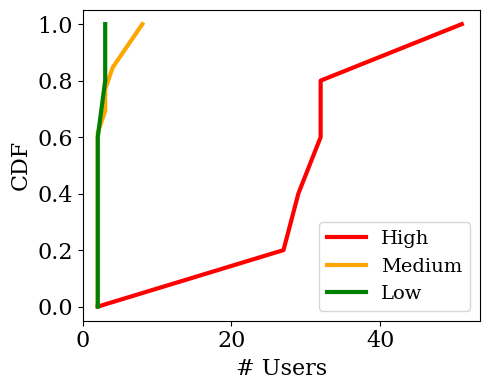}
          \caption{\# Users/community}
            \label{fig:comm_size}
        \end{subfigure}
        \begin{subfigure}[t]{0.48\linewidth}
            \includegraphics[width=\linewidth]{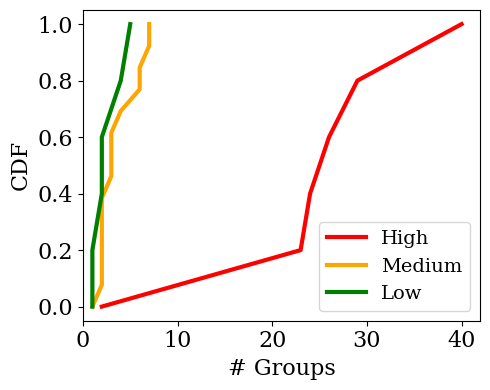}
            \caption{\# Groups/community}
            \label{fig:comm_groups}
        \end{subfigure}
        \caption{Content sharing patterns of WhatsApp communities (Results for week 3).}
        \label{fig:comm_results1}
    \end{center}
\end{figure*}
 
 
 Figure \ref{fig:comm_results1}  shows results on sharing patterns  for communities in each category detected in week 3. Results for the other weeks are similar. 
As shown in Figure  \ref{fig:comm_misinf}, communities in the high category greatly distinguish themselves from the rest   by the   number of messages with misinformation shared (88\% of all misinformation in the week). Indeed, as observed for groups, the greater engagement in misinformation spread tends to be accompanied by more information sharing in general ($\rho$=0.80), more community members ($\rho$=0.67), and  greater group coverage ($\rho$=0.64), as shown in Figures \ref{fig:comm_messages}, \ref{fig:comm_size} and \ref{fig:comm_groups}.  As an example, a very large community with 130 members was responsible for 46\% of all misinformation shared in the week (109 messages), and these messages reached 54 groups, with a  potential audience of 3,289 users (63\% of all users in the network).  

\begin{figure*}[!ttt]
\centering
    \begin{center}
        \begin{subfigure}[t]{0.48\linewidth}
          \includegraphics[width=\linewidth]{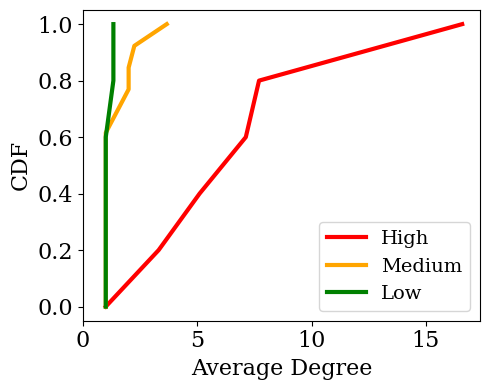}
          \caption{Node degree}
            \label{fig:comm_degree}
        \end{subfigure}
        \begin{subfigure}[t]{0.48\linewidth}
            \includegraphics[width=\linewidth]{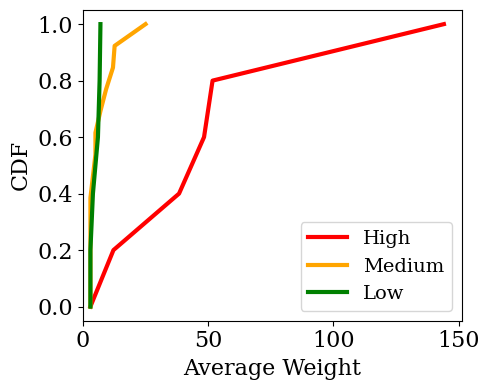}
            \caption{Total edge weight}
            \label{fig:comm_weight}
        \end{subfigure}
        
        \begin{subfigure}[t]{0.48\linewidth}
          \includegraphics[width=\linewidth]{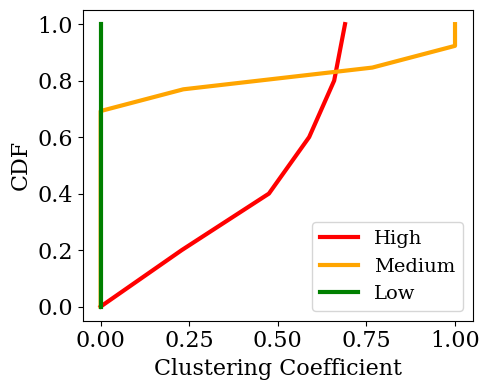}
          \caption{Clustering coefficient}
            \label{fig:comm_clustering_coeff}
        \end{subfigure}
        \begin{subfigure}[t]{0.48\linewidth}
            \includegraphics[width=\linewidth]{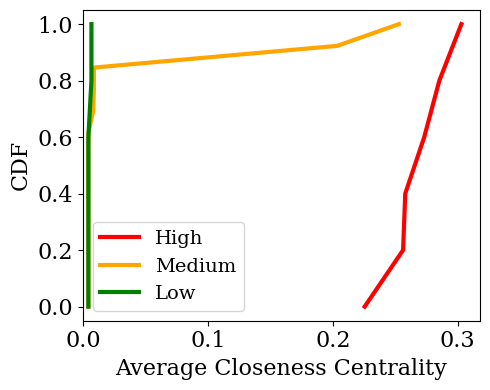}
            \caption{Closeness}
            \label{fig:comm_closeness}
        \end{subfigure}
        \vspace{-0.2cm}
        \caption{Topological properties of communities in the co-sharing networks (Average results per community for week 3).}
        \label{fig:comm_results2}
    \end{center}
\end{figure*}

We now turn to the topological properties of community members. Figure \ref{fig:comm_results2} shows results for communities in each category for week 3. Once again, these results correspond to average values across all community members and are similar to those  obtained for all other weeks. As shown, communities responsible for more misinformation sharing tend to be more strongly connected and well structured, with larger node degrees (Figure \ref{fig:comm_degree}) and larger sum of all edge weights attached to a node (Figure \ref{fig:comm_weight}). Also,   communities in the high category (as some communities in the medium category) tend to have very large average clustering coefficients\footnote{ The clustering coefficient captures the degree to which nodes   tend to cluster together.},  suggesting very strongly connected structures (Figure \ref{fig:comm_clustering_coeff}). Finally, as observed for users and groups, we find that communities more often engaged in misinformation spread tend to agglomerate users who are more centrally located in the co-sharing network (Figure \ref{fig:comm_closeness}), illustrating, once again, the important role that this network has for content dissemination at scale.

\subsection{Temporal Dynamics}


As we argued in \cite{nobre:2020}, the co-sharing networks are very dynamic: around only 30\% of the users persist in the network backbone over time.  Also, community membership changes frequently, although those users who more often share content  tend to remain  in the same community over successive weeks.  Inspired by these prior results, we here offer a final analysis of the dynamics of user participation in misinformation spread.  We focus on individual users and groups since, as just mentioned, community membership changes frequently, thus being hard (even impossible, in some cases) to track the same community across weeks. 

\begin{table*}[tt!]
\centering
\footnotesize\setlength{\tabcolsep}{4.5pt}
\caption{Persistence of users and groups engaged in misinformation over consecutive weeks.}
\label{tab:user_persistence}
\begin{tabular}{|l|c|c|c|c|c|}
\cline{1-6}
 {} & \multicolumn{5}{|c|}{\textbf{Weeks}} \\ 
\cline{2-6} 
\multicolumn{1}{|l|}{\textbf{Persistence }} & 2 & \textbf{3} & 4 & 5 & \textbf{6} \\
\cline{1-6} 
\multicolumn{1}{|l|}{All users engaged in misinformation sharing} & 0.30 & 0.33 &  0.25 & 0.21 & 0.32 \\ 
\multicolumn{1}{|l|}{Top 10 users engaged in misinformation sharing } & 0.1 & 0.2 &  0.1 & 0.4 & 0.1 \\ 
\multicolumn{1}{|l|}{Top 50  users engaged in misinformation sharing} & 0.18 & 0.24 & 0.22 & 0.24 & 0.24 \\
\multicolumn{1}{|l|}{All groups engaged in misinformation sharing} & 0.87 & 0.87 & 0.79 & 0.72 & 0.90 \\ 
\multicolumn{1}{|l|}{Groups in high category} & 0.74 & 0.85 & 0.77 & 0.73 & 0.72 \\ 
\cline{1-6}
\end{tabular}
\end{table*}

Specifically, we consider different sets of users/groups and analyze the dynamics of these sets over successive weeks. 
As in \cite{Ferreira:2020}, we define the persistence of elements in set $A$ from week $w-1$ to week $w$ as the fraction of elements in $A$ at week $w-1$ that remain in the set in week $w$. 
Table \ref{tab:user_persistence} shows the persistence of different sets of users and groups, defined based on their engagement in misinformation sharing. 
As shown, the  sets of all users who shared some misinformation in consecutive weeks have often very small intersections (at most 33\%).  The same can be said if we consider only the most active disseminators of this type of content. For example, often only one or two users remain among the top-10 most active in misinformation spread, and only at most 12 remain among the top-50 ones,  over consecutive weeks.

In contrast, the persistence of groups engaged in misinformation (all groups as well as those in the high category) is quite large. This is partially because, as shown in Table \ref{tab:group_misinf}, some misinformation is shared in most groups in all weeks. But, more than that, we find that, despite the great dynamics of individual behavior,  those groups where most misinformation are shared (category high) remain, to a great extent, the same over consecutive weeks.

\begin{figure*}[!ttt]
\centering
    \begin{center}
        \begin{subfigure}[t]{0.48\linewidth}
          \includegraphics[width=\linewidth]{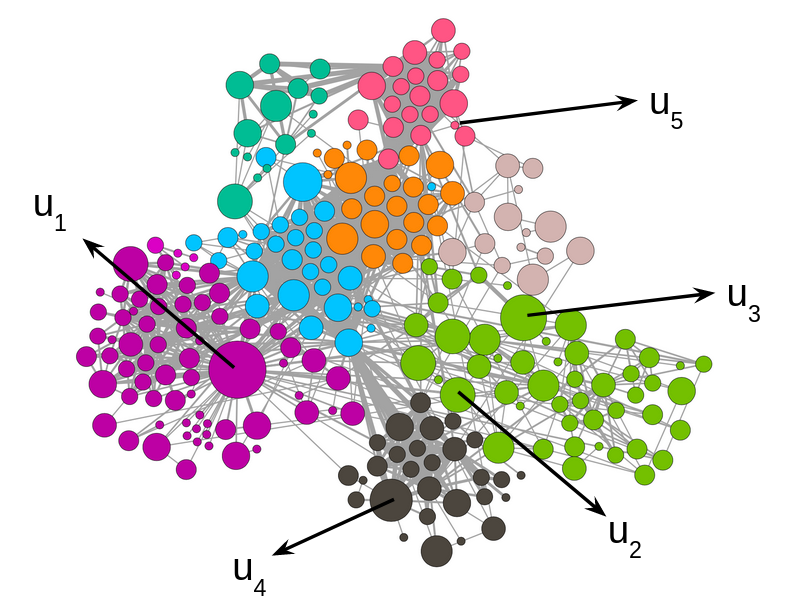}
          \caption{Week 2}
            \label{fig:week2}
        \end{subfigure}
        \hfill
        \begin{subfigure}[t]{0.48\linewidth}
          \includegraphics[width=\linewidth]{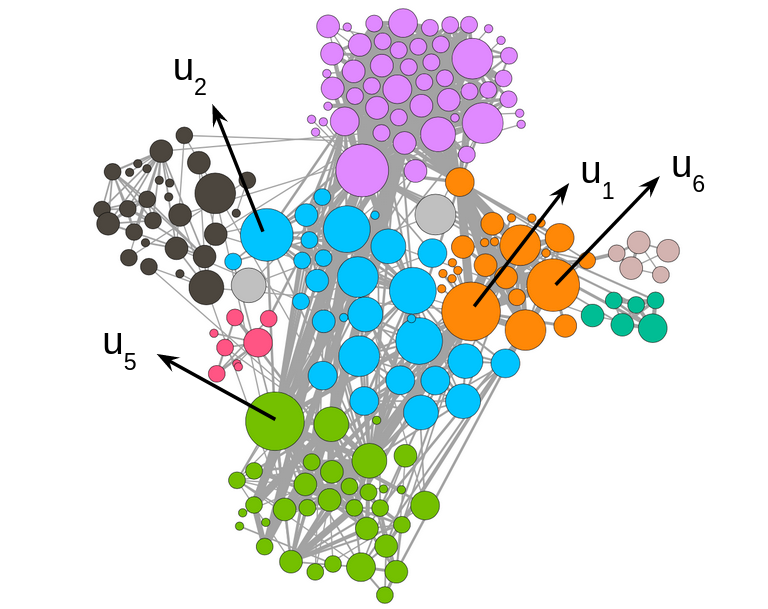}
          \caption{Week 3}
            \label{fig:week3}
        \end{subfigure}
    \end{center}
    \caption{Largest connected component of the  network backbones in weeks 2 and 3 (Node color represents community and node size is proportional to volume of misinformation shared). }
    \label{fig:network_temporal}
\end{figure*}

To illustrate this dynamic, Figure \ref{fig:network_temporal} shows samples the network backbones extracted for weeks 2 and 3. To improve readability, only the largest connected component of each backbone is shown. Node color indicates the community it belongs to\footnote{Although similar colors are used in both components, there is no correspondence between communities in different weeks.}, and node size is proportional to the volume of misinformation shared in the given week.  We select six nodes ($u_1, u_2, ..., u_n$ in the figure) and track them across the two weeks to exemplify the dynamics of their participation in misinformation spread over the two weeks. 

Starting with $u_1$ and $u_2$, we see that both users remain as important sources of misinformation during both weeks (even considering the whole network).  In contrast, nodes $u_3$ and $u_4$ played a key role in misinformation spread in week 2, but disappeared from the (complete) backbones in week $3$. Node $u_5$ is an example of a user who, despite already in the backbone  shared {\it no}  misinformation in week 2 but became one of the top sources of this type of content in the following week. Finally,  node $u_6$ illustrates the case of a newcomer, who shared no misinformation in week 2, but  joined  the backbone  already as one of the top sources of misinformation in week 3.  In sum, we observe that user participation in misinformation spread often changes drastically over time, both in volume and in co-sharing patterns (which affect user's topological properties in the co-sharing network), making it hard to track them even on a weekly basis.

\section{Conclusions and Future Work}\label{conclusion}

We have presented  a hierarchical characterization of user participation in misinformation spread in WhatsApp, with a particular focus on the potential  role of the media co-sharing networks  as a means to gain content visibility and scale of dissemination.  Our analysis  relied on  dataset of messages shared  in publicly accessible WhatsApp groups in Brazil, containing a number of textual, image and audio messages  priorly identified as fake content.\\ 

\noindent {\bf Take-home messages:} 
Our study revealed a number of interesting observations which can be summarized in the following main insights:

\begin{itemize}

\item {\it Despite the great diversity in the participation in misinformation spread across users, groups and communities, there are always a few top sources  of misinformation during each week.} For instance, 
a few users are responsible for a large fraction of all misinformation shared as well as to introduce diversity of misinformation  in the system.  Similarly, a small number of  groups and communities often are responsible for the vast majority of all misinformation shared in the network. Moreover, these groups and communities often have many members who, despite not being engaged in misinformation, are exposed to this type of content by their peers.

\item {\it The underlying media co-sharing network favors misinformation spread at large on   WhatsApp, and both network backbone and network periphery have key roles in such process. }   The network backbone, which emphasizes consistent, strong and potentially coordinated  co-sharing patterns, often  includes  top misinformation sources, being responsible for a great part of all misinformation spread in the platform, with  a clear dominance in terms of content diversity. The periphery, in turn, includes the majority of  all users who shared some misinformation (though to a much lesser degree) who most probably act independently or with no clear evidence of coordinated effort.
  
\item {\it Users forming the backbone tend to build well-structured communities, which may suggest coordination to spread content}. While our prior work \cite{nobre:2020} indicated the existence of user communities that favor content spread, here we characterize them in terms of misinformation. Indeed, our results indicate that such communities manifest a potential coordination effort for misinformation dissemination in the WhatsApp platform.

\item {\it The users, groups and communities more often engaged in misinformation tend to have higher centrality in the network.} By  establishing  a  larger  number  of  stronger  connections in the network, these users, groups and communities end up occupying more central positions in the network, which ultimately may favor misinformation spread.   
Such characteristic was also priorly observed on Twitter \cite{Cheng:2021}, where malicious users have a tendency to occupy more central positions in the information dissemination network. We here broaden the analysis to look beyond individual users, including also WhatsApp groups and communities.

\item {\it Users engaged in misinformation vary over time but remain mostly in the same groups.} Indeed, we noticed that the users more often engaged in misinformation spread change drastically from week to week. Yet, despite the great dynamics of individual behavior,  these users are often members of the same groups.

\end{itemize}

These insights are novel and greatly complement the current literature on 
WhatsApp and closely similar platforms (e.g., Telegram). As already discussed at the end of Section \ref{related},  compared to  prior studies, we here take a completely different perspective,  offering a novel network-driven characterization of user participation in misinformation spread at three different levels of aggregation that constitute inherent organizational components of the WhatsApp platform. By looking at the problem from this novel perspective, we are able to offer a new look into misinformation spread on a widely used application which constitutes one of the main means of communication  in many countries of the world. 
In sum, our work brings new understanding about  misinformation spread on WhatsApp, which, we hope, can drive the future design of more  effective solutions to mitigate the problem, as we further discuss below. \\

\noindent {\bf Potential limitations:} Yet, these insights should be considered in light of the existing limitations that constrain our study, as listed next. 

\begin{itemize} 
\item  {\it Our study covers a relatively short period of analysis.} As already argued, the six-week period considered is of particular interest to study misinformation spread on WhatsApp. Nevertheless, is is arguably too short to capture fundamental properties of users, groups and communities of WhatsApp. A longer period of analyses would be useful to reveal consistent observations as well as potential differences in user behavior depending on   topics of discussion (e.g., politics) and contextual information (e.g., specific events occurring in real world).

\item {\it Reasonably small number of (distinct) messages containing misinformation.} Our dataset includes a few hundred distinct messages containing misinformation each week. This number is inevitably limited by the labeling effort, which is quite costly and requires professional knowledge. We here employed an automatic approach by relying on previously checked facts reported by a number of 
respectable fact checking agencies in Brazil. However, we may be missing some misinformation present in the dataset. Increasing the number of fact checking agencies could help catching more fake content. Yet, the agencies already considered represent some of the most important ones in the country, thus we believe we are already capturing most messages containing misinformation in the dataset, or at least those that had the greatest impact on users, motivating action from these agencies.  Alternatively, extending the number of groups monitored and the period of analysis could lead to a larger amount of messages with misinformation. Similarly, exploring  misinformation spread in video messages, which was not covered in this study, could also open a new front of exploration, by delving into the use of deep fakes \cite{Tolosana:2020, Vaccari:2020}.

\item {\it Focus on the ``potential" channels of information propagation.}
Our dataset does not allow us to closely track the propagation of individual messages. As such, the media co-sharing network offers a representation of the potential channels through which information can be propagated but we cannot state if or when these channels were indeed used. Alternatively, one could  explore the ``reply" feature  to build  cascade models of the propagation of individual messages, similarly to prior studies on other systems \cite{Guo:2018}.  By doing so, one could extend the focus from individual messages to  {\it conversations} triggered by misinformation content \cite{caetano:2019}.


\item {\it Conservative strategy to identify near duplicate content on texts, audios and videos.} For textual messages, we used two standard techniques: a TF-IDF vectorial representation to identify misinformation content and Jaccard similarity to identify near duplicate content. In the future, we could benefit from applying more sophisticated techniques that exploit word and sentence embeddings \cite{mikolov:2013, reimers-gurevych:2019}.
For audio and video messages, we relied on the identification strategy employed by WhatsApp during data transfer to flag duplicate content. By doing so, we may be underestimating the  duplicates, and ultimately building more conservative networks than the real ones. More sophisticated techniques \cite{jegou:2011, kordopatis-zilos:2019} could be employed to catch similar  content, leading to more realistic networks.

\item {\it Study restricted to  WhatsApp.} The current study focus on misinformation on WhatsApp, which is currently the main means of online communication in many countries\footnote{\url{https://www.statista.com/statistics/289778/countries-with-the-most-facebook-users/}}.  Future studies are necessary to assess whether similar results are observed also in other comparable platforms (e.g., Telegram).

\end{itemize}


Our investigation can be extended in several directions. In addition to the many opportunities  of follow-up efforts listed above, a direction of great interest is the design of effective methods to mitigate the spread of misinformation, by leveraging the intrinsic properties of the message sharing patterns, notably properties of the networks interconnecting users. From a practical point of view, and as indicated by our results, focusing efforts on coordinated actions most probably will hit those users who are more actively involved in spreading misinformation. Yet, a large number of less active disseminators, often acting independently, will most probably continue to spread misinformation in the system. Thus, possibly different countermeasures targeting different user profiles should be employed.

\section*{Acknowledgements}
This work has been financed by the Conselho Nacional de Desenvolvimento Científico e Tecnológico (CNPQ), Coordenação de Aperfeiçoamento de Pessoal de Nível Superior - Brasil (CAPES) and Fundação de Amparo à Pesquisa do Estado de Minas Gerais (FAPEMIG).

\bibliography{references.bib}

\appendix
\section{Identification of Messages with Misinformation}
\label{appendix}

For the sake of completeness, we here describe the steps taken in \cite{resende:2019:WWW, resende:2019:WebSci,alexandre:2020} to identify the presence of misinformation in textual, image and audio messages collected from WhatsApp groups. The collection of messages analyzed in these prior studies are the same as those used in this work. Thus, we strictly followed the approaches detailed by the authors. In common, they relied on a set of facts previously checked as {\it fake}, collected from six very popular fact checking agencies in Brazil: “Aos fatos”, “Me engana que
eu posto”, “e-farsas”, “é ou não é (G1)”, “Lupa” and “Boatos.org"
\footnote{ aosfatos.org, veja.abril.com.br/blog/me-engana-que-eu-posto/, www.e-farsas.com, g1.globo.com/e-ou-nao-e/, piaui.folha.uol.com.br/lupa/, and www.boatos.org, respectively}.

For textual content,  WhatsApp messages and checked facts were first pre-processed to remove stopwords and accents as well as to perform stemming. Then, each message/fact was represented by a bag of words, in a TF-IDF vectorial representation. A series of pairwise comparisons between the vectorial representations  of each WhatsApp message and each fact-checked fake content was performed using cosine similarity. Any  message
whose cosine similarity with any of the fake facts was
above a pre-defined threshold was considered suspicious of carrying misinformation. This threshold was chosen as 0.4 after the authors manually evaluated  a sample of message-fact pairs.  All
suspicious messages were then manually   compared
against the fake fact by the authors and, in case of matching, flagged as misinformation \cite{resende:2019:WebSci}.

For the audio messages, first the content was transcripted using Google Cloud's Speech-to-Text 
API\footnote{https://cloud.google.com/speech-to-text/}. A quality check was carried out to keep only messages with higher transcription quality (as estimated by the tool). As done for textual messages, a series of pairwise comparisons between transcripted (audio) messages and collected fake facts were carried out to flag suspicious messages which were then manually analyzed (i.e., the audios were listened by a volunteer) before the decision to label an audio as misinformation \cite{alexandre:2020}.

Finally,  two approaches were applied in conjunction to identify misinformation among images. First, a sample of images was presented to one of the most important fact checking agencies in Brazil -- Lupa -- which identified some of them as misinformation. Second, an automatic approach was employed to expand the set of labeled images. Specifically, the Google Image Search was used to search  the web for each image present in the WhatsApp dataset. 
Given the search results for an image, it was checked whether any of
the returned pages belong to one of the selected six fact-checking domains. 
If so, the fact-checking page was parsed and  the
image was automatically labeled as fake if it was tagged as such in the fact checking webpage \cite{resende:2019:WWW}.

\end{document}